\documentclass[preprints,article,submit,pdftex,moreauthors]{Definitions/mdpi} 

\firstpage{1} 
\pubvolume{1}
\issuenum{1}
\articlenumber{0}
\pubyear{2024}
\copyrightyear{2024}
\datereceived{ } 
\daterevised{ } % Comment out if no revised date
\dateaccepted{ } 
\datepublished{ } 
\hreflink{https://doi.org/} % If needed use \linebreak
\Title{Energy performance of LR-FHSS: analysis and evaluation}

\TitleCitation{Energy performance of LR-FHSS: analysis and evaluation}

\Author{Roger Sanchez-Vital *\orcidA{}, Lluís Casals \orcidB{}, Bartomeu Heer-Salva \orcidF{}, Rafael Vidal \orcidC{}, Carles Gomez \orcidD{} and Eduard Garcia-Villegas \orcidE{}}

\AuthorNames{Roger Sanchez-Vital, Lluís Casals, Bartomeu Heer-Salva, Rafael Vidal, Carles Gomez and Eduard Garcia-Villegas}

\AuthorCitation{Sanchez-Vital, R.; Casals, L.; Heer-Salva, B.; Vidal, R.; Gomez, C.; Garcia-Villegas, E.}
\address[1]{%
Department of Network Engineering, Universitat Politècnica de Catalunya, C/Esteve Terradas, 7, 08860 Castelldefels, Spain; lluis.casals@upc.edu (L.C.); bheersalva@gmail.com (B.H.); rafael.vidal@upc.edu (R.V.); carles.gomez@upc.edu (C.G.); eduardo.garcia@upc.edu (E.G.)}
\corres{Correspondence: roger.sanchez.vital@upc.edu}

\abstract{Long-range frequency hopping spread spectrum (LR-FHSS) is a pivotal advancement in the LoRaWAN protocol that is designed to enhance the network's capacity and robustness, particularly in densely populated environments. Although energy consumption is paramount in LoRaWAN-based end devices, {this is the first study} in the literature, to our knowledge, that models the impact of this novel mechanism on energy consumption. In this article, we provide a comprehensive energy consumption analytical model of LR-FHSS, focusing on three critical metrics: average current consumption, battery lifetime, and energy efficiency of data transmission. The model is based on measurements performed on real hardware in a fully operational LR-FHSS network. While in our evaluation, LR-FHSS can show worse consumption figures than LoRa, we find that with optimal configuration, the battery lifetime of LR-FHSS end devices can reach 2.5 years for a 50 min notification period. For the most energy-efficient payload size, this lifespan can be extended to a theoretical maximum of up to 16 years with a one-day notification interval using a cell-coin battery.}


\begin{document}

\section{Introduction}
In the last decade, low-power wide Area networks (LPWANs) have emerged as a family of long-range, low-power communication technologies suitable for many Internet of Things (IoT) applications~\cite{MEKKI20191,fi12030046,rfc8376}.

LoRaWAN has arisen as one of the most popular LPWAN technologies, with~around one billion devices predicted to use this technology in the near future~\cite{survey_lorawan_technology,survey_lorawan_iot,hofmann2021comparison}. In~order to expand the capabilities of LoRaWAN, the~LoRa Alliance recently introduced a new physical layer called long-range frequency hopping spread spectrum (LR-FHSS) \cite{lr-fhss_overview}. Exploiting techniques such as intrapacket fragmentation, frequency diversity, and~increased transmission redundancy, LR-FHSS is expected to enable network deployments with greater node density, robustness, and~coverage~\cite{lr-fhss_overview,DoS_scenario_lrfhss,outage_probability_lrfhss,outage_probability_lrfhss_DoS,transceiver_DoS_lrfhss,uplink_transmission_policies_DoS,dfh_lrfhss,jung2024lrfhsstransceiverdirecttosatelliteiot}.

One particularly promising use case for LR-FHSS is direct-to-satellite IoT (DtS-IoT), which is a~field with significant momentum~\cite{lr-fhss_overview,DoS_scenario_lrfhss,outage_probability_lrfhss,outage_probability_lrfhss_DoS,transceiver_DoS_lrfhss,uplink_transmission_policies_DoS,dfh_lrfhss,jung2024lrfhsstransceiverdirecttosatelliteiot,DoS_survey,iot_rate_splitting,ris_satellite_relay}. DtS-IoT provides communication means for IoT devices in remote areas, where terrestrial network infrastructure may not be feasible or practical to deploy. In~DtS-IoT, when an IoT device is visited by a satellite, the~former can transmit frames (e.g., carrying sensed data) to the latter, which can act as a~gateway.

Since the power grid is not available for many IoT devices in general, and~especially for those using LR-FHSS, such devices often need to rely on a {limited energy source, such as a simple battery or energy harvesting (based on either natural sources or wireless transmission systems~\cite{IRS_SWIPT})}. Therefore, although~energy efficiency is not the main objective of the mechanism, determining the energy performance of LR-FHSS is crucial. However, to~our knowledge, {the present paper is the first study that} has specifically addressed this~topic.

In this paper, we provide a detailed analytical model of the current consumption of an LR-FHSS IoT device (end device, in~LoRaWAN terminology), derived from measurements on real hardware within a complete LoRaWAN network (i.e., including an end device, gateway, and~network server) supporting LR-FHSS. This represents the first such model in the academic literature. We also use the model to determine the lifetime of a battery-operated device and the energy efficiency of data transmission with LR-FHSS. In~addition, we compare the energy performance of LR-FHSS with that of classic LoRaWAN physical layer alternatives. The~evaluation results present trade-offs that depend on the data rate (DR), operational mode, and payload sizes for~every performance metric. Among~other findings, the results show that battery lifetime can approach 2.5 years with a 50 min notification interval when utilizing the proper configuration and can even reach a theoretical maximum of up to 16 years with a more infrequent interval of 1 day between messages for~the most energy-efficient packet~size.

The remainder of the paper is organized as follows. Section~\ref{sec:relatedwork} reviews related work. Section~\ref{sec:overview} provides background concepts on LoRaWAN and LR-FHSS. Section~\ref{sec:analytical_model} presents our model, which is used in Section~\ref{sec:evaluation} to evaluate and discuss the current consumption, battery lifetime, and energy efficiency of an LR-FHSS end device. Finally, Section~\ref{sec:conclusions} concludes the~paper.

%%%%%%%%%%%%%%%%%%%%%%%%%%%%%%%%%%%%%%%%%%
\section{Related~Work}\label{sec:relatedwork}
This section provides an overview of the literature related to LR-FHSS, with~a particular focus on energy consumption. As~aforementioned, energy consumption is a critical feature in IoT. For~LoRaWAN, some detailed energy consumption models have been published~\cite{energy_consumption_model_lora_lorawan,lorawan_energy_performance,refined_energy_consumption_model_lorawan}. However, {there is only one contribution that models LR-FHSS energy consumption, and it only focuses on uplink transmission (i.e., it does not model the complete LoRaWAN transmission procedure) \cite{mikhaylov_lrfhss_consumption}.}

The body of work on LR-FHSS has recently increased~\cite{lr-fhss_overview,DoS_scenario_lrfhss,outage_probability_lrfhss,outage_probability_lrfhss_DoS,transceiver_DoS_lrfhss,uplink_transmission_policies_DoS,dfh_lrfhss,jung2024lrfhsstransceiverdirecttosatelliteiot,lrfhss_real_world_packet_traces,maldonado2024enhancinglrfhssscalabilityadvanced}. Most of the studies on LR-FHSS predominantly investigate its known strengths---namely, coverage and scalability---when applied to its main use case, DtS-IoT~\cite{DoS_scenario_lrfhss,outage_probability_lrfhss,outage_probability_lrfhss_DoS,transceiver_DoS_lrfhss}. Only three published works give some attention to the energy consumption performance of LR-FHSS~\cite{uplink_transmission_policies_DoS,dfh_lrfhss,semtech2021lrfhss}.

Using a custom-made simulator, the authors of~\cite{uplink_transmission_policies_DoS} {concluded} that LR-FHSS can improve the deployment scalability by a factor of 75x at the expense of 30\% higher device power consumption compared to the legacy LoRa modulation. However, this comparison is only based on the transmission time-on-air of the different physical layer approaches considered, which misses several significant contributions to energy consumption (see Section~\ref{sec:analytical_model}).  In~another work~\cite{dfh_lrfhss}, the~impact of the frequency hopping sequence (FHS) in LR-FHSS {was} studied using a LoRaWAN Class B end device based on commercial transceivers and an SDR-based gateway. For~a very specific setup and two FHS proposals, energy efficiency results {were} provided, with~an increase of 5.20 times for the proposed dynamic frequency hopping (DFH) scheme compared to a tailor-made transmit power control method. However, it is not clear how the authors {determined} the energy consumption to derive those results. Finally, a~technical report released by the LR-FHSS chip manufacturer presents the characteristics of the technology and provides several performance figures, including the power consumption and battery lifetime of an LR-FHSS end device~\cite{semtech2021lrfhss}. However, the~results {were} obtained by using a limited set of states based on current consumption values of unknown origin, and~the model used to produce the results is not~given.

{The authors of~\cite{mikhaylov_lrfhss_consumption} provided a study on energy consumption of LR-FHSS. However, they only modeled LR-FHSS in the uplink and did not account for a complete LoRaWAN-based transmission procedure; they also included acknowledged and unacknowledged procedures. Furthermore, in~the present paper, we also study the energy efficiency of LoRaWAN with LR-FHSS and compare its performance with that of LoRaWAN using the legacy LoRa PHY.}

Therefore, to~the best of our knowledge, the~present paper is the first to provide a detailed energy consumption model that allows for the prediction of the current consumption, the~battery lifetime, and~the energy efficiency of an LR-FHSS~end device.

%%%%%%%%%%%%%%%%%%%%%%%%%%%%%%%%%%%%%%%%%%
\section{Overview of LoRaWAN and~LR-FHSS}\label{sec:overview}
Created by the LoRa Alliance~\cite{lora_alliance}, the~LoRaWAN protocol defines the media access control (MAC) layer and provides extensive networking capabilities to enable long-range, low-power communication on top of several physical layer (PHY) alternatives. The~original PHY used in LoRaWAN, known as LoRa, was developed by Semtech, a~founding member of the alliance. This section presents the main LoRaWAN protocol concepts and the LoRa and LR-FHSS underlying PHY~layers.

\subsection{LoRaWAN}\label{sec:LoRaWAN}
LoRaWAN networks are composed of end devices (EDs), gateways (GWs), and a network server (NS). To~enable communication between EDs and GWs, the~former use pure ALOHA as the medium access mechanism and~one of the PHY modulations allowed by the protocol—LoRa, frequency shift keying (FSK), or~LR-FHSS—to transmit frames. After~the frames are demodulated by the GW, they are forwarded to the NS, typically over an IP backhaul. The~LoRaWAN specification defines the communication protocol and system architecture~\cite{lorawan_specification_1_0_4}, while the Regional Parameters specification tailors LoRaWAN networks to operate efficiently and in compliance with regional regulatory requirements~\cite{regional_parameters_v104}.

\begin{figure}[H]
%\begin{center}
\includegraphics[width=1\textwidth]{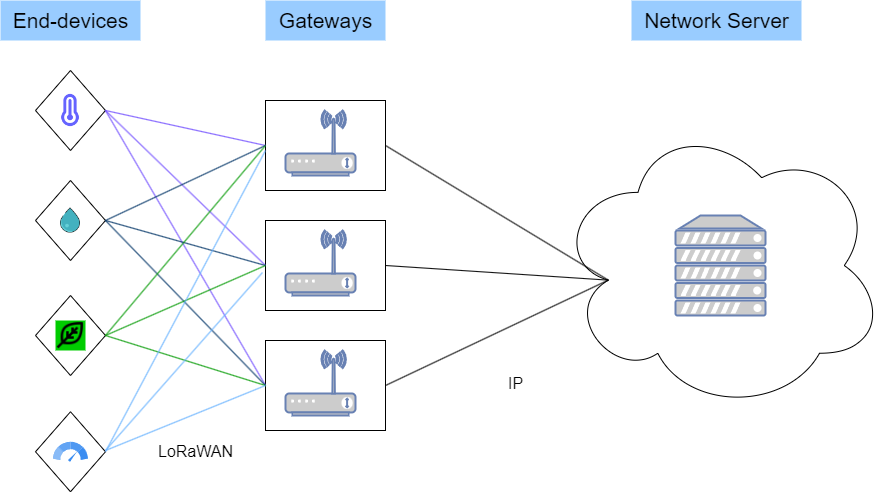}
\caption{LoRaWAN network~architecture.}
\label{fig:lorawanArch}
%\end{center}
\end{figure}

The topology of LoRaWAN networks is a star-of-stars. EDs use one or more GWs to transmit uplink frames to the NS, as~shown in Figure~\ref{fig:lorawanArch}. However, the~NS replies to an ED via downlink messages through a single~GW.

To accommodate diverse application needs, LoRaWAN supports three different classes of EDs: Class A, Class B, and~Class~C.

\begin{figure}[H]
\includegraphics[width=0.7\textwidth]{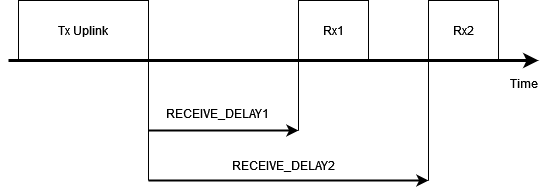}
\caption{Class A operation with~one uplink transmission followed by two receive~windows.}
\label{fig:rxwindows}
\end{figure}

\begin{itemize}
\item \textbf{Class A}: %MDPI: Please confirm if the bold is unnecessary and can be removed. The following highlights are the same. %Authors: If it is possible, we find it necessary to have the items in bold so the text is more easily readable
This is the default operational mode for LoRaWAN networks, and~all devices must support it. Class A devices by default remain in a sleep state and perform uplink transmissions asynchronously when needed. There are two downlink (or receive) windows that follow each uplink frame. These windows are used to receive commands or data from the NS. The~parameters RECEIVE\_DELAY1 and RECEIVE\_DELAY2 specify the time between the end of the uplink transmission and the start of the first and second receive windows, respectively. Figure~\ref{fig:rxwindows} presents a diagram of this behavior, where {the specification recommends a default value for RECEIVE\_DELAY1 of 1 s} and RECEIVE\_DELAY1 + 1 s for RECEIVE\_DELAY2. Downlink transmission can only happen after an uplink frame: a~mechanism that enhances energy efficiency but~limits the device's %Please ensure meaning has been retained.
applicability. For~the EU868 region, the~LoRaWAN standard specifies that the DR used for the first receive window should match that of the corresponding uplink frame, with~an additional parameter called Rx1DROffset, %MDPI: Please confirm if the italics is unnecessary and can be removed. %Authors: Indeed, we can remove it (already done).
which can range from 0 to 5. In~our study, we set this value to 0, which is the default in \mbox{the specification.}
\item \textbf{Class B}: In addition to Class A operation, Class B devices can schedule additional receive slots. The~key characteristic is that they allow for more predictable and regular opportunities for the NS to send downlink messages to the devices. Because~of this, Class B devices can receive commands or downlink data independently of uplink traffic. Support for Class B is optional.
\item \textbf{Class C}: Except for when they are transmitting, Class C devices offer practically continuous receive windows. Although~it is the most power-consuming out of the three classes, it is suitable for equipment that runs on-grid, and~it minimizes downlink latency. Support for Class C is likewise optional.
\end{itemize}

In this article, we base our study on Class A devices, as~they are the most popular and are used in the most energy-constrained~applications.

Transmission via LoRaWAN can be set to be confirmed or unconfirmed. When reliable transmission is required, the~ED transmits an uplink frame and then waits for a confirmation from the NS during a receive window. On~the other hand, in~the unconfirmed mode, the~ED is unaware of whether the data were correctly received because there is no confirmation frame from the~NS.

\begin{figure}[H]
\begin{center}
\includegraphics[width=0.99\textwidth]{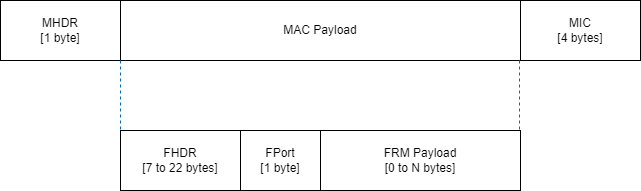}
\caption{LoRaWAN MAC frame~structure.}
\label{fig:loraWAN-MAC}
\end{center}
\end{figure}

Figure~\ref{fig:loraWAN-MAC} shows the MAC frame structure in LoRaWAN. The~first field, the~one-byte sized MAC header (MHDR), identifies the type of transmission. Three components make up the MAC payload: the frame payload (FRM Payload), the~frame header (FHDR), and~the FPort. The~FPort is only present when data are carried via the FRM Payload, which has a one-byte size limit. The~FOpts field, which has a size of 0 (no operations included) to 15 bytes, is part of the FHDR, which has a size that ranges from 7 to 22 bytes. The~message integrity code (MIC), the~final field in the MAC frame, enables the verification of frame~integrity.

LoRaWAN defines the concept of DR, which indicates a PHY layer mechanism and a particular radio interface configuration leading to a specific PHY layer bit rate. For~each DR, there is a maximum payload size specified. The~maximum MAC payload size and the maximum application PDU (FRM Payload) size, presented as M and N values, respectively, when FOpts is not present are shown in Table~\ref{tab:loraMaxPayl}.

\begin{table}[H]
\caption{Maximum MAC payload sizes for LoRaWAN in column M and~maximum application PDU size (FRM Payload) in column N for~each DR and without FOpts~field.}
%\begin{center}
\begin{tabularx}{1\textwidth}{CCC}
\toprule
\textbf{DR} & \textbf{M (bytes)} & \textbf{N (bytes)} \\ \midrule
0 & 59 & 51 \\
\midrule
1 & 59 & 51 \\
\midrule
2 & 59 & 51\\
\midrule
3 & 123 & 115\\
\midrule
4 & 250 & 242\\
\midrule
5 & 250 & 242\\
\midrule
6 & 250 & 242\\
\midrule
7 & 250 & 242\\
\midrule
8 & 58 & 50 \\
\midrule
9 & 123 & 115 \\
\midrule
10 & 58 & 50 \\
\midrule
11 & 123 & 115 \\
\midrule
12--15 & \multicolumn{2}{c}{Not Defined} \\
\midrule
\end{tabularx}
\label{tab:loraMaxPayl}
%\end{center}
\end{table}
\unskip

\subsection{LoRa}\label{sec:lora}
LoRa is a PHY radio communication technology that was released in 2009. It employs a modulation technique known as chirp spread spectrum (CSS), which uses frequency-modulated chirps to encode data. This method can accomplish long-distance communications and robustness against interference. In~CSS, every symbol is represented by a unique series of chirps. The~frequency is swept linearly across a predetermined bandwidth, either upwards or downwards, to~produce these chirps. The~symbol value determines the initial frequency of each chirp. The~key parameters to take into account in LoRa are:

\begin{enumerate}
\item \textbf{Bandwidth (BW):} BW is the frequency range that the chirp signal covers. The~data rate tends to increase with bandwidth. In~the EU region, there are two possible BW values: 250 kHz and 125 kHz.
\item \textbf{Spreading factor (SF):} The SF is the quantity of chirps required to represent one symbol. Each symbol is translated into \(2^{SF}\) chirps. Bit rate and range are impacted by the SF; larger SFs lead to longer on-air times and lower bit rates, but~they also boost sensitivity and range. Particularly, when used in LoRaWAN, the~supported LoRa SFs range from 7 to~12.
\item \textbf{DR:} DR depends on the combination of distinct SF and BW values. The~LoRaWAN Regional Parameters specification defines a range of possible DR values that vary depending on the region~\cite{regional_parameters_v104}. There are eight potential DRs in the EU region, with~the DR index ranging from 0 to 7. However, the~latter can only be used with FSK modulation. Table~\ref{tab:DRlora} summarizes the possible configurations and shows the resulting physical bit rate for each one. For~LR-FHSS, which we will address in Section~\ref{sec:lrfhss}, DRs 8 to 11 were~added.
\item \textbf{Coding rate (CR):} In order to prevent bit corruption, forward error correction (FEC) is used, with different possible CR values being between~0 and 4. These correspond to coding rates from 4/5 to 4/8, with CR = 0 (or coding rate 4/5) being the default value. For~LR-FHSS, coding rate values 1/3 and 2/3 have been introduced.
\end{enumerate}\vspace{-6pt}

%Authors: %Authors: If it is possible, we find it necessary to have the items in bold so the text is more easily readable

\begin{table}[H]
\caption{LoRa EU863-870 DR~characteristics.}
%\begin{center}
\begin{tabularx}{1\textwidth}{m{2cm}<{\centering}CCC}
\toprule
{\bf DR Index} & {\bf Modulation} & {\bf Configuration} & {\bf Physical Bit Rate [bps]}  \\ \midrule
0        & LoRa    & SF12/125 kHz    & 250  \\
1        & LoRa    & SF11/125 kHz    & 440 \\
2        & LoRa    & SF10/125 kHz    & 980 \\
3        & LoRa    & SF9/125 kHz    & 1760 \\
4        & LoRa    & SF8/125 kHz    & 3125 \\
5        & LoRa    & SF7/125 kHz    & 5470 \\
6        & LoRa    & SF7/250 kHz    & 11,000 \\
7        & FSK     & 50 kbps      & 50,000 \\ \bottomrule
\end{tabularx}
\label{tab:DRlora}
%\end{center}
\end{table}

The LoRaWAN specification also defines {support for the adaptive data rate (ADR) mechanism}, which allows the NS to evaluate the link quality and adapt the EDs to the channel conditions for optimal performance by~modifying the DR, transmission power, and maximum number of retransmissions as appropriate. %Please ensure meaning has been retained.

\subsection{LR-FHSS}\label{sec:lrfhss}
\textls[-15]{LR-FHSS is an extension of the LoRa physical radio modulation~\cite{regional_parameters_v104}. The~purpose of this new PHY technique is to improve data transmission in congested networks wherein capacity is reduced by duty cycle restrictions, channel availability, and~collision probability. LR-FHSS uses frequency hopping spread spectrum (FHSS), which also improves the link range and~allows numerous devices to communicate on the same operating channel simultaneously while their signals can still be appropriately received and demodulated by the GW. {LR-FHSS is the LoRa Alliance's implementation of FHSS, which is a~technique used in other wireless technologies like Bluetooth or early Wi-Fi amendments. FHSS involves rapidly switching a signal's carrier frequency across a predetermined set of channels, making it more robust against interference and unauthorized access.} These features are especially helpful in satellite networks, where EDs and GWs are far away from each other and there is high node density. LR-FHSS is only used to perform uplink transmission; downlink transmission continues to employ LoRa PHY. There are four LR-FHSS DRs: DR8 to DR11, and all of them are included.}  %Please ensure meaning has been retained.

With LR-FHSS, the~ED splits the payload into pieces and sends each piece over a separate physical channel, which is explained later. To~provide redundancy, PHY headers are sent three times for DR8/DR10 and twice for DR9/DR11. Moreover, there are differences in the coding rates: DR8/DR10 use a CR of 1/3, whilst DR9/DR11 use a CR of 2/3. {This means that DR8/DR10 are more robust than DR9/DR11, but~at the expense of a lower bit rate. A~more robust coding scheme might be helpful in low-signal-quality conditions (i.e., low SINR), but~the longer transmission time could also increase the chance of collisions. This could lead to worse performance with respect to DR9/DR11, especially in scenarios with a large number of EDs~\cite{DoS_scenario_lrfhss}.} The LR-FHSS bitrates are 162 bps (for DR8/DR10) and \mbox{325 bps} (for DR9/DR11).

\begin{figure}[H]
\includegraphics[width=0.99\textwidth]{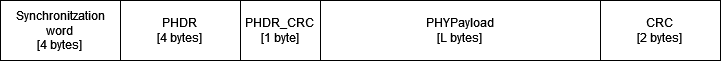}
\caption{LR-FHSS PHY frame~structure.}
\label{fig:lrfhss-msg}
\end{figure}

The LR-FHSS PHY frame structure is composed of the header and the payload (referred to as PHYPayload), %AE unsure if different usage is intentional or should be standardized between 'FRM Payload' and 'PHYPayload'
as~shown in Figure~\ref{fig:lrfhss-msg}. The~header information gives the GW the tools it needs to reassemble the payload from the ED. It contains the channel hopping sequence, payload length, DR, number of header replicas, and~coding rate. As~we previously mentioned, the~header is communicated more than once for redundancy; one header is transmitted at a fixed rate for a duration of 233.472 ms. The~variable-sized payload, which is divided into segments with a duration of 102.4 ms each, comes after the header. After~a 233.472 ms header segment and~after a 102.4 ms payload segment, there is a frequency channel hop. A~two-byte payload CRC is the last component of the PHY frame~structure.

The operating channel width (OCW) for DR8/DR9 and DR10/DR11 is 137 kHz and 336 kHz, respectively. Each OCW is divided into several occupied bandwidths (OBWs) of 488 Hz. The~minimum separation between physical channels, or~the grid, %Please ensure meaning has been retained.
is 3.9 kHz, which implies a separation of eight OBWs. Up~to 688 OBW physical channels can be present in a 336~kHz OCW, but~due to the 3.9 kHz separation constraint, only 86 physical channels can be used for each uplink frame transmission (336  kHz/3.9 kHz %MDPI: We removed the italics of units. Please confirm this revision. %Authors: OK.
$\simeq$ 86). This results in eight grids or~groupings of physical channels ($8 \times 86 = 688$). On~the other hand, for~an OCW of 137~kHz, there are up to 280 OBW physical channels, but~only 35 can be used due to the grid restrictions. The~ED first selects the OCW and physical channel at random before the start of an uplink transmission. Subsequent physical channels are chosen in a pseudo-random fashion, which guarantees an even carrier distribution. These parameters are summarized in Table~\ref{tab:lrfhss-params} for the EU863-870 band~\cite{lr-fhss_overview}.

\begin{table}[H]
\caption{LR-FHSS PHY parameters for region~EU863-870.}%MDPI: 1. We removed the vertical lines. Please confirm. 2. Please confirm if this table has no table header. %Authors: We confirm it has no table header.
\centering
\begin{tabularx}{\textwidth}{Xcccc}
\toprule
{DR index}   & 8     & 9     & 10    & 11  \\ \midrule
Operating channel width (OCW) [kHz]%MDPI: Please confirm if the bold is unnecessary and can be removed. The following highlights are the same.  %Authors: It can be removed (already did so).
&\multicolumn{2}{c}{137}  & \multicolumn{2}{c}{336} \\ \midrule
Occupied bandwidth (OBW) [Hz]       & \multicolumn{4}{c}{488} \\ \midrule
Minimum separation between hopping channels (grid) [kHz]       & \multicolumn{4}{c}{3.9} \\ \midrule
Number of usable physical channels per LR-FHSS transmission   & \multicolumn{2}{c}{35}   & \multicolumn{2}{c}{86}   \\ \midrule
Available physical channels for frequency hopping     & \multicolumn{2}{c}{280 (8 $\times$ 35)}   & \multicolumn{2}{c}{688 (8 $\times$ 86)}  \\ \midrule
Coding rate (CR)   & 1/3   & 2/3   & 1/3   & 2/3  \\ \midrule
{Physical bit rate [bps]}     & 162   & 325   & 162   & 325 \\ \bottomrule
\end{tabularx}
\label{tab:lrfhss-params}
\end{table}

In contrast to LoRa channels, as~long as LR-FHSS packets stay within the designated bandwidth of the GW, they can be demodulated. Prior knowledge of certain frequencies or channel hopping sequences is not required. This enables several transmitters to operate simultaneously with distinct channel hopping sequences if~the GW is able to listen to the whole channel bandwidth at the same time~\cite{lr-fhss_overview}. In~contrast to original LoRaWAN use cases, this enables for the simultaneous reception of hundreds of packets, which increases the complexity of signal recognition at the receiver but makes it appropriate for networks with high device density (e.g., satellite-scale).

%%%%%%%%%%%%%%%%%%%%%%%%%%%%%%%%%%%%%%%%%%
\section{Current Consumption Model of an LR-FHSS~ED}\label{sec:analytical_model}
In this section, we present analytical models for crucial energy performance parameters of an LR-FHSS ED, such as average current consumption, battery lifetime, and~energy cost of data transmission. We assume that the ED transmits application data periodically, which emulates the behavior of many sensors. We also assume Class A LoRaWAN operation, considering that its support is mandatory for LoRaWAN equipment and~it is also the most popular in LoRaWAN due to its energy efficiency compared to the other classes~\cite{survey_lorawan_technology}. This section is organized into three subsections. The~first one describes the experimental scenario that we have used in order to perform the current consumption measurements our model is based on. The~other two subsections provide the LR-FHSS ED current consumption model along with the battery lifetime and energy cost of data transmission for~confirmed and unconfirmed transmission, respectively.

\subsection{Experimental~Scenario}\label{sec:experimentalScenario}
\begin{figure}[H]
\includegraphics[width=\textwidth]{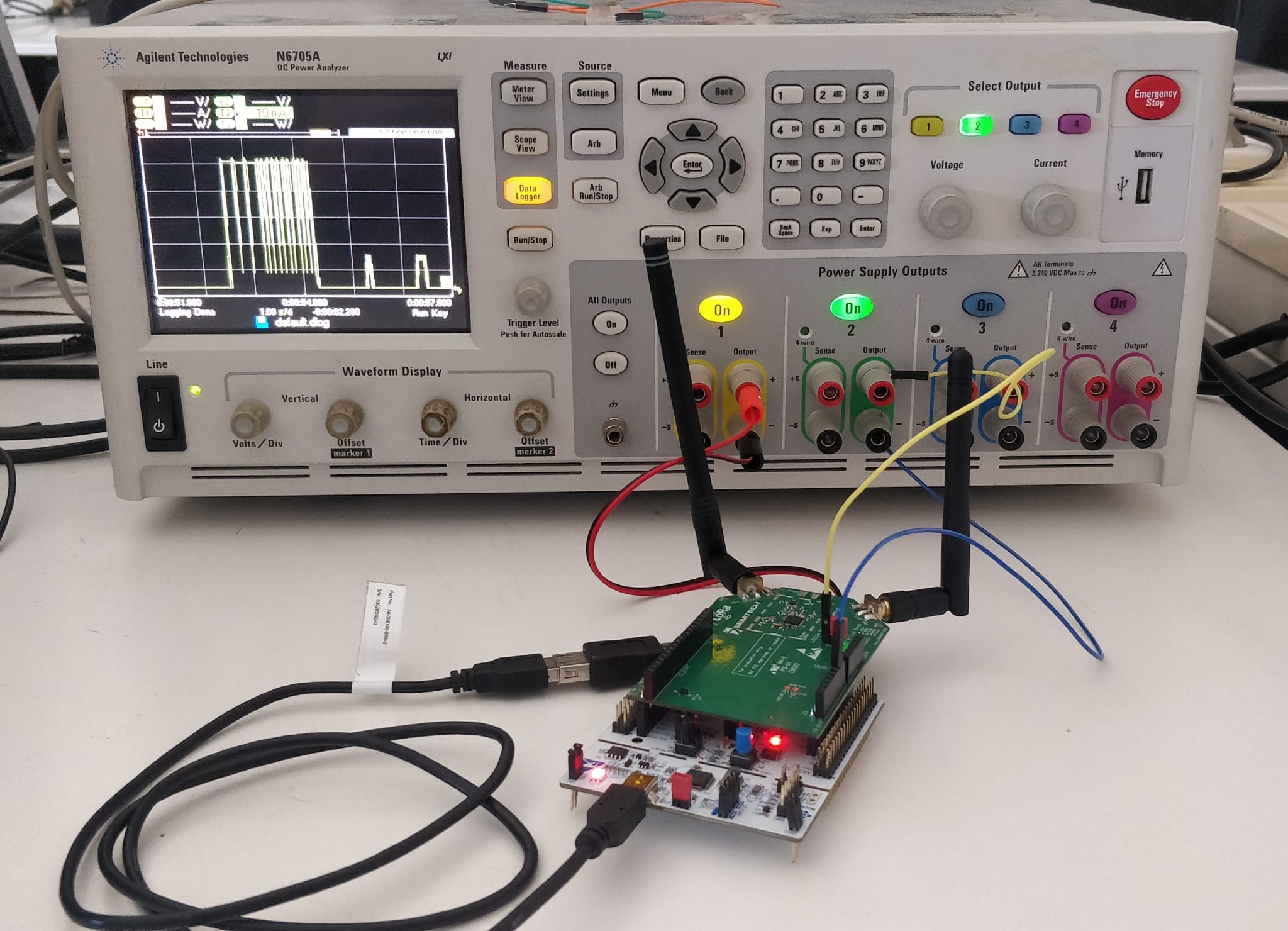}
\caption{Experimental %MDPI: Please check this image, it cannot correctly dispay in the generated PDF. %Authors: I see it properly when I generate the PDF.
scenario for measuring the current consumption of the considered LR-FHSS~ED.\label{fig:scenario}}
\end{figure}

The experimental scenario for which we carry out the measurements is a complete LoRaWAN network. To the best of our best knowledge, this is the first time in the academic literature that a complete LoRaWAN network---comprising an ED, GW, and NS, all supporting LR-FHSS---has been tested in this manner. %Please ensure meaning has been retained.
For the ED, we use a LR1121DVK1TBKS development kit from Semtech~\cite{semtech_lr1121dvk1tbks}, which is composed of a Nucleo L476 board alongside a LR-FHSS-capable radio interface based on the LR1121 chipset~\cite{semtech_lr1121}. For~the GW, we use a Kerlink Wirnet iBTS Compact, which is LR-FHSS-compatible after a firmware update~\cite{kerlink_wirnet_ibts}. We deploy an instance of Chirpstack version 4.8.1 on-premise to~act as the NS~\cite{chirpstack}. To~perform the measurements, we use the Keysight N6705A DC power analyzer~\cite{keysight_n6705a}, which includes two power supply outputs (cf. Figure~\ref{fig:scenario}). In~this scenario, the~ED is supplied 5 V via the USB interface at one of the outputs to feed the Nucleo L476 board, and~the energy consumption of the radio interface is measured through a 3.3 V second power supply that only feeds this part of the system.
In the measurements, the~transmit power of the ED is configured to +14 dBm, which is the maximum value for the band in the EU868 region~\cite{regional_parameters_v104}. {This is the transmit power that would be used to achieve the longest link range. Therefore, our work allows us to determine the energy performance in scenarios where an end device is configured to achieve the longest link range. In~this study, as~the link range does not affect the measurements,} both the ED and the GW are located in an indoor environment and are at~a distance of around 2 m from each~other.
\vspace{3pt}
\unskip

\subsection{Unconfirmed~Transmission}

Initially, we aim to model the average current consumption of an LR-FHSS ED in the unconfirmed mode, denoted $I_{avg\_unTx}$. To~this end, we first identify and characterize the different states the ED goes through to perform an unconfirmed transmission in~terms of the duration and current consumption for each state. To~create a realistic model of the ED's behavior, we carry out measurements using the experimental scenario presented in Section~\ref{sec:experimentalScenario}. The~measurements are performed only on the radio module of the ED. This is because our ED hardware platform comprises components (e.g., LEDs, communication interfaces, etc.) that are useful for development but unnecessarily increase energy consumption compared with that of a production-environment~ED.

We assume that the LoRaWAN ED transmits data units periodically; therefore, we model its current consumption over one period. Each period includes the transmission of a frame (along with the necessary related LoRaWAN protocol procedures), with~the device remaining in a sleep state~otherwise.

\begin{figure}[H]
\begin{adjustwidth}{-\extralength}{0cm}
\centering
\includegraphics[width=17.5cm]{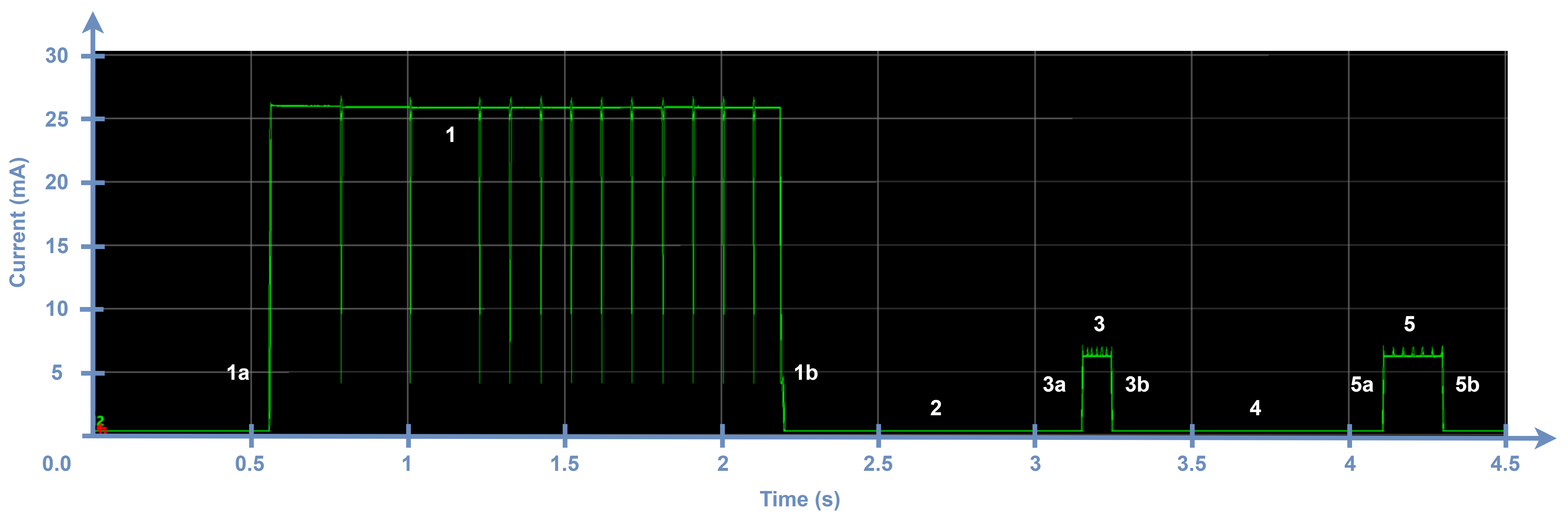}
\end{adjustwidth}
\caption{Current %MDPI: Please confirm if the explanation of the red arrow needs to be added in the figure caption. %Authors: Which red arrow?
consumption profile of ED transmitting an unconfirmed uplink frame with DR8. The~PHYPayload size is 17~bytes.}
\label{fig:noACkuplinkDR8}
\end{figure}

An unconfirmed transmission comprises one uplink frame transmission and two subsequent receive windows. {Recall that LR-FHSS is only used in uplink transmission; therefore, downlink traffic employs LoRa PHY.} Figure~\ref{fig:noACkuplinkDR8} shows a power analyzer capture of the complete procedure for the transmission of an unconfirmed data unit using DR8 with a PHYPayload size of 17 bytes (i.e., 4 bytes of FRM Payload size). We next identify and characterize each state involved in the transmission procedure (labeled with a tag composed of a number and a letter).
\vspace{-9pt}

Details for every state depicted in Figure~\ref{fig:noACkuplinkDR8} are shown in Table~\ref{tab:statesTxnoACK}. These states correspond to different operations of the radio interface. We measured the duration and current consumption of each state for several individual transmission processes, and~the differences we found were~negligible.

\begin{figure}[H]
\includegraphics[width=0.7\textwidth]{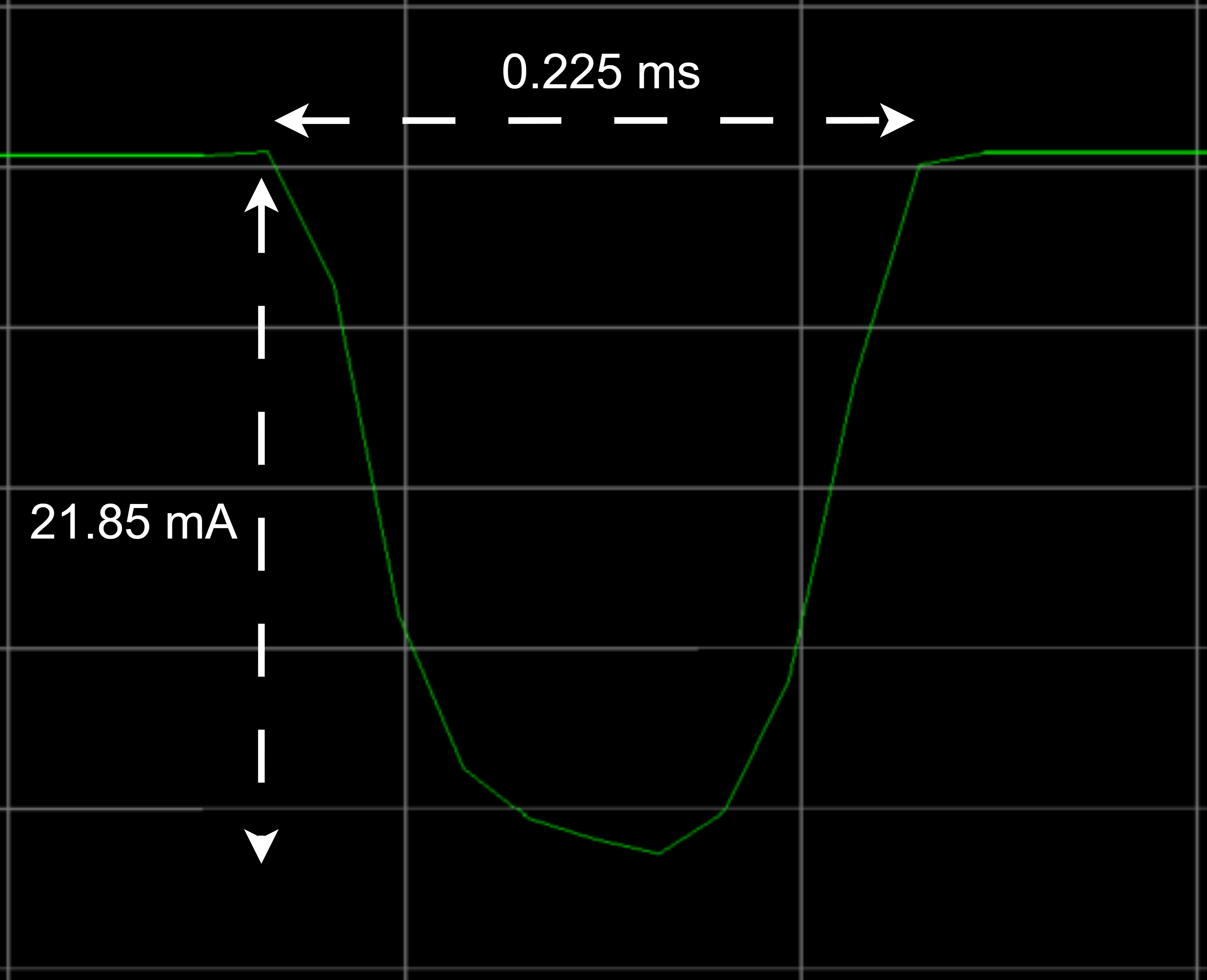}
\caption{Expanded view of a frequency channel hop in an uplink transmission with~LR-FHSS.}
\label{fig:channelGapDR8}
\end{figure}

The actual frame transmission happens in state 1, whereas the first and the second receive windows correspond to states 3 and 5, respectively. States 1a, 3a, and~5a are the initial states whereby the radio module is preparing for the subsequent main state. States 1b, 3b, and~5b, on~the other hand, correspond to the post-operational stages, during~which the radio module transitions to the sleep state. The~radio interface remains in sleep mode in states 2, 4, and~6. Moreover, taking a closer look at state 1, we observe the impact of the LR-FHSS frequency channel hops (i.e., physical carrier hops) on current consumption. Figure~\ref{fig:channelGapDR8} shows an expanded view of one such frequency hop. Each hop implies a brief and smooth drop in current~consumption.

The average current consumption in the unconfirmed mode, $I_{avg\_unTx}$, is modeled in Equation~(\ref{eq:IavgNoACK}). $T_{Period}$ denotes the period between two consecutive transmissions. $T_j$ and $I_j$ represent the duration and current consumption, respectively, of a specific state $j$ in Table~\ref{tab:statesTxnoACK}. Note that frequency channel hops are encompassed in state 1.
\begin{equation}\label{eq:IavgNoACK}
I_{avg\_unTx} = \frac{1}{T_{Period}}\sum _{j}T_j \times I_j
\end{equation}

Then, the~duration of the sleep interval, $T_{Sleep}$, which depends on several variables like $T_{Period}$, is shown in Equation~(\ref{eq:tsleep}).
\begin{equation}\label{eq:tsleep}
T_{Sleep} = T_{Period} - T_{act\_unTx}
\end{equation}

\noindent where $T_{act\_unTx}$ represents the sum of all non-sleep-state durations of the states involved in the transmission of a frame (see Equation~(\ref{eq:TactnoACK})).
\begin{equation}\label{eq:TactnoACK}
\begin{gathered}
T_{act\_unTx} = T_{preTx} + T_{Tx} + T_{postTx} + T_{Rx1wait} + T_{preRx1} +  \\
+ T_{Rx1} + T_{postRx1} + T_{Rx2wait} + T_{preRx2} + T_{Rx2} + T_{postRx2}
\end{gathered}
\end{equation}

\vspace{-9pt}

The total transmission time of a frame, $T_{Tx}$, depends on the time needed to transmit the physical header replicas, the~physical layer payload, and~the total time of the frequency channel hops performed during such header and payload transmission, denoted $T_{header}$, $T_{payload}$, and $T_{freqHops}$, respectively, as~expressed in Equation~(\ref{eq:TTx}).
\begin{equation}\label{eq:TTx}
T_{Tx} = T_{header} + T_{payload} + T_{freqHops}
\end{equation}

The duration of the header transmission, $T_{header}$, can be obtained as defined in Equation~(\ref{eq:theader}) \cite{regional_parameters_v104}. $N$ indicates the number of times that the header is transmitted, the values of which are listed in Table~\ref{tab:equationValues}. Note that the CR value refers to the coding rate used for the payload transmission, not for the header transmission. For~the latter, CR is always defined as $CR = 1/2$.
\begin{equation}\label{eq:theader}
T_{header} = N \times 233.472 \; ms
\end{equation}

$T_{payload}$ can be calculated as expressed in Equation~(\ref{eq:tpayload}). $L_{PHY}$ and $M$ denote the PHYPayload size and the fragment size, respectively, with both being measured in bytes. The~possible values for $M$ are shown in Table~\ref{tab:equationValues}. Note that other works based on the expressions %Please ensure meaning has been retained.
listed in Table~108 %MDPI: Please check if it's from ref.28. %Authors: It is correct.
of the Regional Parameters v1.0.4 specification~\cite{regional_parameters_v104} provide a different equation for $T_{payload}$ (see Appendix \ref{sec:appendix}), which we found is not accurate. We encountered inconsistencies between the measured values and the theoretical calculations predicted in the expressions provided in the current version of the specification~\cite{regional_parameters_v104}. Following discussions with members of the LoRa Alliance, we provide the correct equation for $T_{payload}$ as Equation~(\ref{eq:tpayload}). We need to take into account that the total amount of bytes to be transmitted in order to send the PHYPayload is $L_{PHY}$ + 2 + 6/8, since in addition to $L_{PHY}$, the 2-byte CRC and 6 bits of Trellis termination also need to be sent. FEC uses a convolutional encoding, so to help the decoder, six zeros are pushed in the encoder at the end of the packet. \begin{equation}\label{eq:tpayload}
T_{payload} = \frac{L_{PHY}+2+\frac{6}{8}}{M} \times 102.4 \; ms
\end{equation}

\vspace{-9pt}

\begin{table}[H]
\caption{Characterization of the states of an unconfirmed uplink transmission in LR-FHSS with DR8 (indicated in Figure~\ref{fig:noACkuplinkDR8}) in~terms of their time and current~consumption.}
\label{tab:statesTxnoACK}
\begin{adjustwidth}{-\extralength}{0cm}
\begin{tabularx}{\fulllength}{m{1.8cm}<{\centering}Cm{1.8cm}<{\centering}Cm{1.8cm}<{\centering}m{1.8cm}<{\centering}}
\toprule
\multirow{2.4}{*}{\textbf{State Index}} & \multicolumn{1}{c}{\multirow{2.4}{*}{\textbf{Description}}} & \multicolumn{2}{c}{\textbf{Time}} & \multicolumn{2}{c}{\textbf{Current}} \\
\cmidrule{3-6}
&& \textbf{Parameter} & \textbf{Value (ms)} & \textbf{Parameter} & \textbf{Value (mA)} \\
\midrule
1a & Pre-transmission & $T_{preTx}$ & $2.370$ & $I_{preTx}$ & $3.8$ \\

1 & Transmission &  $T_{Tx}$ & Equation~(\ref{eq:TTx}) & $I_{Tx}$ & $25.7$ \\

1' & Channel hop &  $T_{Hop}$ & $0.225$ & $I_{Hop }$ & $12.3$ \\

1b & Post-Transmission & $T_{postTx}$ & See {Table}~\ref{tab:statesDRnoACK} & $I_{postTx}$ & $3.7$ \\

2 & Wait until Rx1 & $T_{Rx1wait}$ & $1000$ & $I_{Rx1wait}$ & $0.0005$ \\

3a & Pre-receive window Rx1 & $T_{preRx1}$ & $1.300$ & $I_{preRx1}$ & $2.3$ \\

3 & Receive window Rx1 & $T_{Rx1}$ & See Table~\ref{tab:statesDRnoACK} and Equation~(\ref{eq:Trx1})& $I_{Rx1}$ & $5.8$ \\

3b & Post-receive window Rx1 & $T_{postRx1}$ & $0.700$ & $I_{postRx1}$ & $1.2$ \\

4 & Wait until Rx2 & $T_{Rx2wait}$  & $911.2$ & $I_{Rx2wait}$ & $0.0005$ \\

5a & Pre-receive window Rx2 & $T_{preRx2}$ & $1.500$ & $I_{preRx2}$ & $1.8$ \\

5 & Receive window Rx2 & $T_{Rx2}$ & $198.4$ & $I_{Rx2}$ & $5.8$ \\

5b & Post-reception window Rx2 & $T_{preRx2}$ & $0.700$ & $I_{postRx2}$  & $1.2$ \\

6 & Sleep state & $T_{Sleep}$ & Equation~(\ref{eq:tsleep}) & $I_{Sleep}$  & $0.0005$ \\
\bottomrule
\end{tabularx}
\end{adjustwidth}
\end{table}
\vspace{-9pt}
\begin{table}[H]
\caption{Values of N and M depending on the CR used for the~payload.\label{tab:equationValues}}
\begin{tabularx}{1\textwidth}{CCC}
\toprule
\textbf{CR}	& \textbf{N}	& \textbf{M (bytes)}\\
\midrule
1/3		& 3			& 2\\
2/3		& 2	    & 4\\
\bottomrule
\end{tabularx}
\end{table}

$T_{hop}$ denotes the duration of a frequency channel hop, referred to as state 1' in Table~\ref{tab:statesTxnoACK}. To~compute the amount of time that the ED spends hopping between channels, we have to first calculate the total number of said hops for a single uplink transmission, denoted as $N_{hops}$ and~calculated using Equation~(\ref{eq:nhops}). Again, this depends on the CR that is used in the transmission of the payload. Note that for~a given uplink transmission there will be one frequency channel hop after each header transmitted and~one channel hop after each fragment transmission (except for the last one).
\begin{equation}\label{eq:nhops}
N_{hops} = N + \left \lfloor \frac{L_{PHY}+2+\frac{6}{8}}{M} \right \rfloor
\end{equation}

Finally, the~actual duration of the total frequency channel hopping time for an uplink frame, $T_{freqHops}$, can be calculated as in Equation~(\ref{eq:thops}).
\begin{equation}\label{eq:thops}
T_{freqHops} = N_{hops} \times T_{hop}
\end{equation}

We next determine $T_{Rx1}$. After~transmission of the uplink frame, the~NS can transmit a downlink frame to the ED: either a data frame or an acknowledgment (ACK). The~downlink frame is intended to be received in one of the two receive windows. Even if the uplink transmission is performed with LR-FHSS, downlink transmission will use the LoRa modulation. A~receive window must be at least as long as the physical layer preamble of the downlink transmission to ensure the ED can detect the incoming downlink frames. The~preamble consists of eight symbols for DR0 and DR1 and~twelve symbols for the rest of the LoRa DRs (i.e., DR3, DR4, and DR5); the symbols are %Please ensure meaning has been retained.
denoted as $N_{symb}$. When using LR-FHSS, the~DR for Rx1 is DR1 for uplink frames transmitted with DR8 or DR10, whereas it is DR2 for uplink frames sent with DR9 or DR11. By~default, the~DR for Rx2 is fixed to~DR0.

Rx1 will always be opened by the ED regardless of the communication mode (i.e., unconfirmed or confirmed). We calculate $T_{Rx1}$ using Equation~(\ref{eq:Trx1}).
\begin{equation}\label{eq:Trx1}
T_{Rx1} = N_{symb} \times T_{symb}
\end{equation}

Following the LoRa specification~\cite{semtech_sx1276}, Equation~(\ref{eq:Tsymb}) models the duration of a symbol, $T_{symb}$, in~the first receive window.
\begin{equation}\label{eq:Tsymb}
T_{symb} = \frac{2^{SF}}{BW}
\end{equation}

However, we have observed that there is a discrepancy between the calculations using Equation~(\ref{eq:Trx1}) and the values measured in our testbed (cf. Figure~\ref{fig:noACkuplinkDR8}). Specifically, we measured a shorter value in our scenario, as~the device waits for six symbols before closing Rx1. Therefore, we can use Equation~(\ref{eq:Trx1}) to calculate the duration of the receive windows, but~by using $N_{symb}$ = 6. In~Rx2, the~ED might be utilizing channel activity detection (CAD), which is a~power-saving technique that shortens the duration of the second receive window when no incoming frame is being detected in that window~\cite{semtech_cad_lora_packets}. Then, $T_{Rx2}$ denotes the duration of Rx2 and is calculated using Equation~(\ref{eq:Trx2}).
\begin{equation}\label{eq:Trx2}
T_{Rx2} = \frac{2^{SF} + 32}{BW}
\end{equation}

To complete the whole set of experiments, Table~\ref{tab:statesDRnoACK} depicts the values missing from Table~\ref{tab:statesTxnoACK}, which were computed via the given equations or measured depending on the DR and FRM Payload size. For~DR0 and DR5, values were extracted or derived from~\cite{lorawan_energy_performance}. We found that $T_{freqHops}$ accounts for approximately 0.2\%, on~average, of~the total transmission~time.

%MDPI: We added border line, please confirm. %Authors: Confirmed.
\begin{table}[H]
%\centering
\caption{Transmission variables depending on the DR and~the FRM Payload size. For~the latter, two cases are considered: the minimum one (i.e., 1 byte) and~the maximum one permitted for each~DR.}
\label{tab:statesDRnoACK}
\begin{tabularx}{\textwidth}{LLCCCC}
\toprule

\multicolumn{1}{l}{}                  &                  & \textbf{DR8/DR10}                & \textbf{DR9/DR11}               & \textbf{DR0}                  & \textbf{DR5}                  \\
\midrule
%\cline{3-6}
\multicolumn{2}{l}{$T_{postTx}$ (ms)}                  & $10.40$                       & $12.40$                      & $0.676$                     & $0.676$                     \\

\multicolumn{2}{l}{$T_{Rx1}$ (ms)}                     & $99.20$                       & $49.50$                      & $198.40$                     & $16.40$                      \\

\multicolumn{2}{l}{$T_{header}$ (ms)}                  & $700.4$                    & $466.9$                   & $401.4$                           & $12.54$                           \\

\multicolumn{2}{l}{FRM Payload Maximum Size (bytes) }   & $50$    & $115$  & $51$  & $242$  \\
\midrule
\multirow{2}{*}{$T_{payload}$ (ms)}   & 1 byte         & $870.4$                      & $435.2$                     & $753.7$                           & $33.79$                           \\

%& 50 bytes    & $3379.2$  & $1689.6$ & $2064.384$      & $105.472$      \\

& $Max$            & $3379.2$                     & $3353.6$                    & $2228.2$                           & $387.1$                           \\
\midrule
\multirow{2}{*}{$T_{freqHops}$ (ms)}  & 1 byte         & $2.475$                        & $1.350$                      & N/A                           & N/A                           \\

%& 50 bytes     & $7.875$   & $4.05$   & N/A      & N/A      \\

& $Max$            & $7.875$                      & $7.650$                     & N/A                           & N/A                           \\
\midrule
\multirow{2}{*}{$T_{Tx}$ (ms)}        & 1 byte           & $1573.3$                    & $903.5$                   & $1056.7$                    & $65.50$                      \\

%& 50 bytes     & $4087.5$ & $2160.6$ & $2781.4$ & $96.51$  \\

& $Max$            & $4087.5$                   & $3828.2$                  & $2793.5$                    & $399.6$                     \\
\bottomrule
\end{tabularx}
\end{table}

We can now utilize $I_{avg\_unTx}$ to calculate the lifetime of a battery-operated LR-FHSS ED that periodically performs unconfirmed transmissions. This performance parameter is essential, as~it gives an approximation of the amount of time that a battery-operated LR-FHSS device may function without requiring the recharging or replacing of its battery. Equation~(\ref{eq:Tlifetime}) shows how $T_{lifetime\_unTx}$ depends on the battery capacity, $C_{battery}$, expressed in mAh, and~on $I_{avg\_unTx}$, expressed in mA.
\begin{equation}\label{eq:Tlifetime}
T_{lifetime\_unTx} = \frac{C_{battery}}{I_{avg\_unTx}}
\end{equation}

Finally, we can also calculate the energy cost of data transmission per bit, $EC_{unTx}$, which refers to the amount of energy consumed by an LR-FHSS ED to transmit one bit of application data, as~shown in Equation~(\ref{eq:energeticCostunACK}). $V$ and $L_{data}$ respectively denote the supply voltage and the application-layer protocol data unit (i.e., the~FRM Payload size).
\begin{equation}\label{eq:energeticCostunACK}
EC_{unTx} = \frac{I_{avg\_unTx} \times V \times T_{Period}}{L_{data}}
\end{equation}

\subsection{Confirmed~Transmission}

In contrast to unconfirmed uplink transmission, in~confirmed uplink transmission, the~NS informs the ED via a confirmation downlink frame that the uplink frame was successfully received. In~this section, we model the current consumption of the ED when it performs a confirmed uplink~transmission.

\begin{figure}[H]
\includegraphics[width=\textwidth]{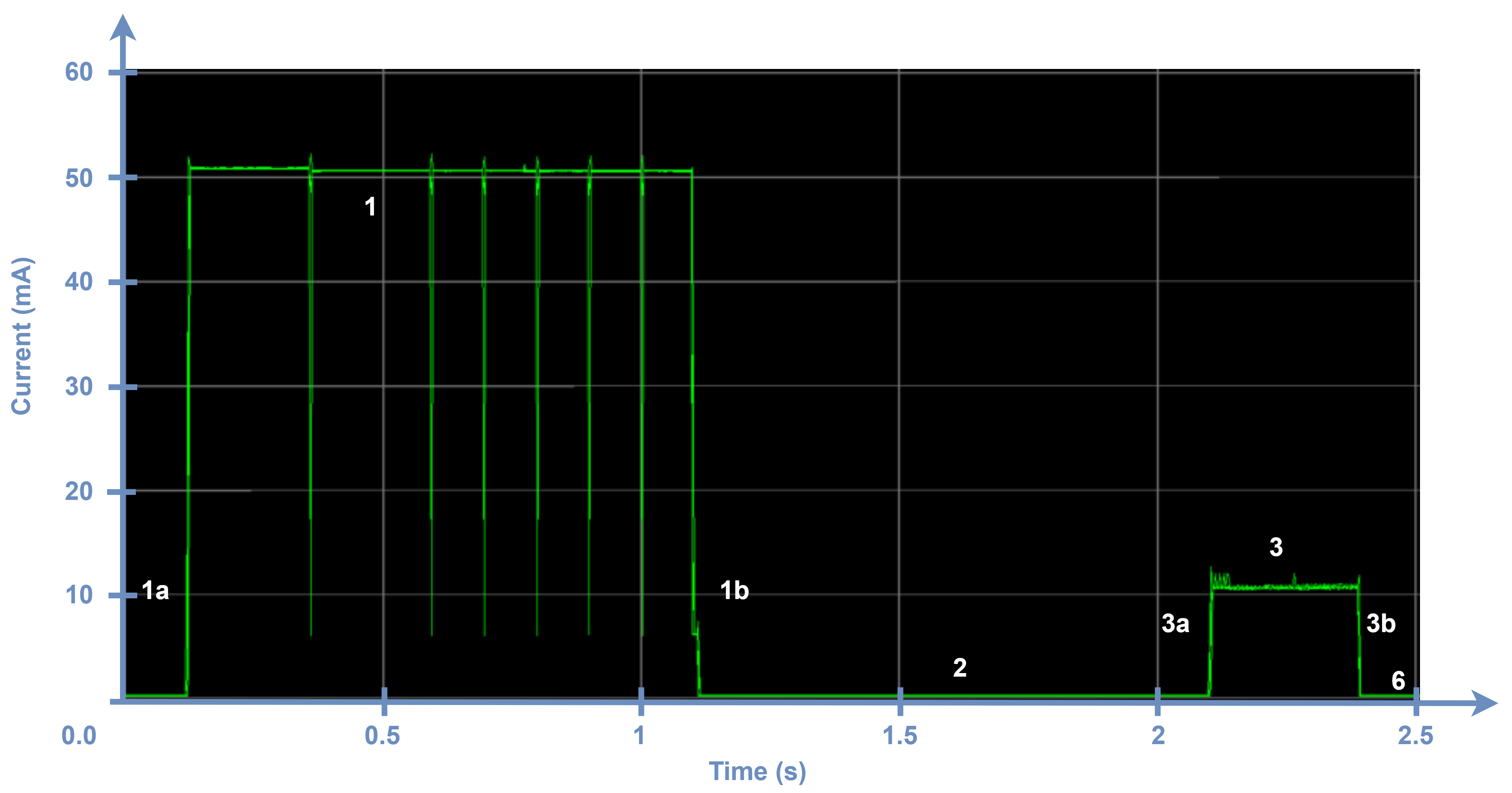}
\caption{Current consumption of an ED during the process of transmitting a confirmed frame with DR9. In~this case, the~ACK is received in the first window (Rx1), so the second window is not~opened.\label{fig:ACkuplinkDR9}}
\end{figure}

\begin{figure}[H]
\begin{adjustwidth}{-\extralength}{0cm}
\includegraphics[width=17.5cm]{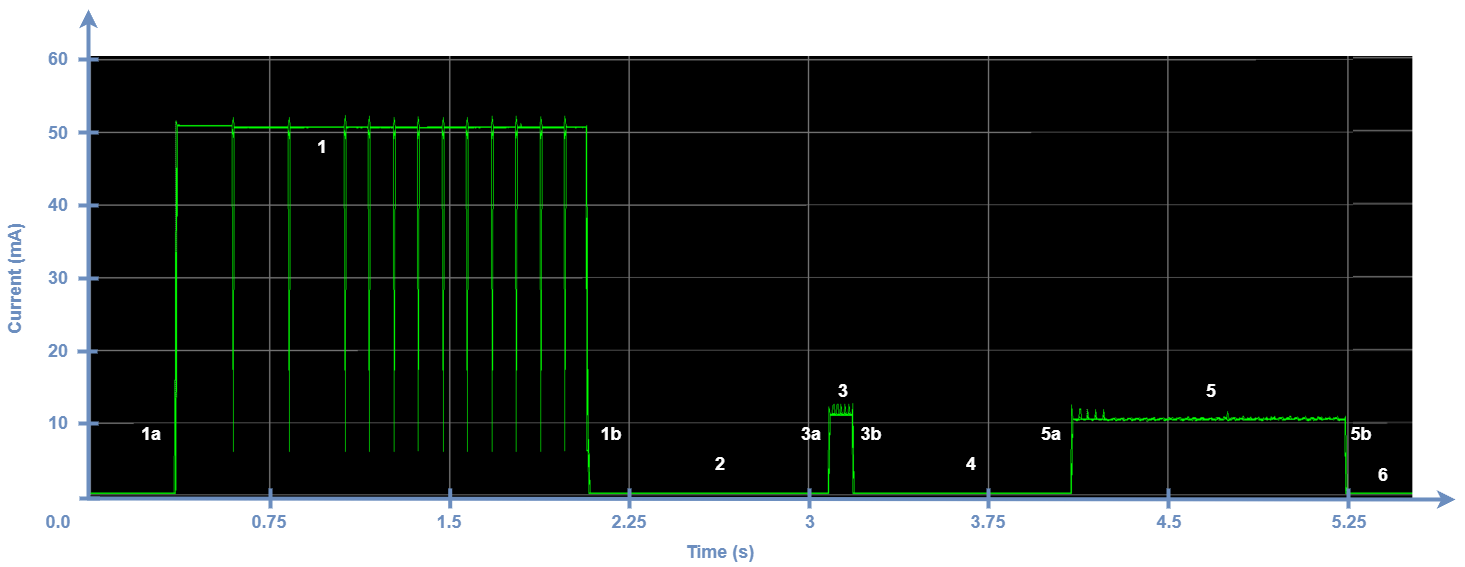}
\end{adjustwidth}
\caption{Current consumption of an ED during the process of transmitting a confirmed frame with DR8. The~ACK is received in the second window (Rx2).\label{fig:ACkuplinkDR8}}
\end{figure}

In a confirmed transmission, the~ACK can be sent in the first window (see Figure~\ref{fig:ACkuplinkDR9}) or in the second one (Figure~\ref{fig:ACkuplinkDR8}) with probabilities of $p_1$ and $p_2$, respectively. Equation~(\ref{eq:IavgACK}) models the average current consumption of an ED performing confirmed transmissions periodically, $I_{avg\_ACKTx}$, where $I_{avg\_ACKTx1}$ and $I_{avg\_ACKTx2}$ indicate the average current consumptions that correspond to receiving the downlink frame in the first and second receive windows, respectively. For~the purpose of the model, we consider that the probability of receiving an ACK in the first or in the second receive window is $p_1 = p_2 = 0.5$.
\begin{equation}\label{eq:IavgACK}
I_{avg\_ACKTx} = p_1\times I_{avg\_ACKTx1} + p_2 \times I_{avg\_ACKTx2}
\end{equation}
\vspace{-20pt}

Figure~\ref{fig:ACkuplinkDR9} illustrates the case where the ACK is received in the first receive window, eliminating the need for the second receive window. This reduction in the number of states consequently leads to lower energy consumption. Specifically, compared with unconfirmed transmission, states 4, 5a, 5, and~5b from Table~\ref{tab:statesTxnoACK} are removed, and~$T_{Rx1}$ is variable depending on the DR and payload length used. To~calculate $I_{avg\_ACKTx1}$, we use Equation~(\ref{eq:IavgNoACK}), considering the states applicable to a confirmed transmission, the~$T_{Rx1}$ value that corresponds to the DR used for the uplink transmission, as~shown in Table~\ref{tab:statesACKRx1}, and~that an ACK has no FRM~Payload.

\vspace{-9pt}

\begin{table}[H]
\caption{States and parameters affected by a confirmed uplink transmission when the ACK occurs in the first receive~window.}%MDPI:  Please confirm that this table has no table header. %Authors: We confirm there is no table header.
\label{tab:statesACKRx1}
\begin{tabularx}{1\textwidth}{CCC}
\toprule \multicolumn{2}{l}{\textbf{State}} %MDPI: Please confirm if the bold is unnecessary and can be removed. The following highlights are the same. %Authors: I would not remove the bold for the first column. If it is all right from your side, I would leave it as I have done now.
& 3      \\ \midrule
\multicolumn{2}{l}{\textbf{Parameter}}                   & $T_{Rx1}$ (ms) \\ \midrule
\multicolumn{1}{l}{\multirow{4}{*}{\textbf{DR}}} & 8/10 & $576.4$         \\ %\cline{2-7}
\multicolumn{1}{l}{}                    & 9/11 & $286.6$       \\ %\cline{2-7}
\multicolumn{1}{l}{}                    & 0 & $991.8$       \\ %\cline{2-7}
\multicolumn{1}{l}{}                    & 5 & $41.20$      \\ \bottomrule
\end{tabularx}
\end{table}

If the ACK is received in the second receive window, the~duration of the latter is extended, as~depicted in Figure~\ref{fig:ACkuplinkDR8}. Therefore, $T_{Rx2}$ will change to a value that depends on the DR used. Typically, the~data rate used for this window is the most robust one (i.e., DR0). From~our measurements, $T_{Rx2} = 1141$  ms. Therefore, $I_{avg\_ACKTx2}$ is equal to $I_{avg\_unTx}$, obtained as in Equation~(\ref{eq:IavgNoACK}) except~for this specific $T_{Rx2}$ value. Current consumption in the first receive window is higher when the ACK is received in it, but~it is lower on average for the whole transmission process because the second receive window is eliminated, as~opposed to unconfirmed uplink transmissions. Therefore, the~current consumption increases when the ACK is received in the second receive window due to the contribution of both receive windows, including the first receive window and its already large duration (by default, it is configured to DR0).

After determining $I_{avg\_ACKTx}$, let $T_{lifetime\_ACKTx}$ denote the lifetime of a battery-operated LR-FHSS ED performing confirmed transmissions periodically. $T_{lifetime\_ACKTx}$ can be calculated as shown in Equation~(\ref{eq:TlifetimeACK}).
\begin{equation}\label{eq:TlifetimeACK}
T_{lifetime\_ACKTx} = \frac{C_{battery}}{I_{avg\_ACKTx}}
\end{equation}

Finally, the~energy cost of transmitting each user data bit for confirmed uplink frames, $EC_{ackTx}$, is modeled in Equation~(\ref{eq:energeticCostACK}).
\begin{equation}\label{eq:energeticCostACK}
EC_{ackTx} = \frac{I_{avg\_ACKTx} \times V \times T_{Period}}{L_{data}}
\end{equation}

%%%%%%%%%%%%%%%%%%%%%%%%%%%%%%%%%%%%%%%%%%
\section{Evaluation}\label{sec:evaluation}
This section evaluates and discusses the three main energy performance metrics that we modeled in the previous section for an LR-FHSS ED: average current consumption, lifetime, and~energy cost of data transmission. The~section is divided into three main parts: one for each performance~metric.

Due to their equal characteristics in terms of maximum payload size and physical layer bit rate, the~following analysis does not differentiate between DR8 and DR10 nor between DR9 and DR11. In~the evaluation, we also include DR0 and DR5, the~slowest and fastest mandatory LoRa DRs, respectively, for~the sake of comparison with LR-FHSS. Duration values for $T_{Tx}$ in these cases are extracted or derived from~\cite{lorawan_energy_performance}.

\subsection{Current~Consumption}

%In order to determine the impact of LR-FHSS on current consumption, we have evaluated this parameter both with and without confirmed transmissions, using Equations~(\ref{eq:IavgNoACK}) and (\ref{eq:IavgACK}). We used the minimum FRM Payload size of 1 byte, the maximum FRM Payload size that is supported by all LoRaWAN DRs (i.e., 50 bytes), and the maximum allowed FRM Payload for LR-FHSS DRs and for DR0 and DR5 of LoRa.

We evaluate the current consumption of an LR-FHSS ED, both with and without confirmed transmissions, using Equations~(\ref{eq:IavgNoACK}) and (\ref{eq:IavgACK}), respectively. %Please ensure meaning has been retained.
We use the minimum FRM Payload size of 1 byte and the maximum allowed FRM Payload size for the LR-FHSS DRs and~for DR0 and DR5 of~LoRa.

As introduced in Section~\ref{sec:overview}, as~the ED works over the EU863-870 MHz band, the~1\% duty cycle limitation must be considered in the evaluation. Since $T_{Tx}$ varies depending on the DR, we provide the minimum interval between consecutive transmissions for each DR (see Table~\ref{tab:dutyCycle})  allowed by the 1\% duty cycle~restriction.

\begin{table}[H]
\caption{Minimum interval between two consecutive transmissions in order to comply with the 1\% duty cycle regulation, depending on the DR and FRM Payload size from Table~\ref{tab:loraMaxPayl}. Values for DR0 and DR5 are obtained from prior work~\cite{lorawan_energy_performance}.}
\label{tab:dutyCycle}
\begin{tabularx}{1\textwidth}{CCC}
\toprule
\multirow{2.5}{*}{\textbf{DR}} & \multicolumn{2}{c}{\textbf{Minimum Frame Transmission Period (s)}}     \\ \cmidrule{2-3} %MDPI: We added bold style for the first row, please confirm. %Authors: OK
& \textbf{1-Byte Payload} & \textbf{Max Payload} \\ \midrule
0                 & $105.7$ & $279.3$ \\
5                & $6.550$  & $39.96$ \\
8/10                  & $157.3$ & $408.7$ \\
9/11                & $90.35$  & $382.8$ \\
\bottomrule
\end{tabularx}
\end{table}
%MDPI: Please confirm if the bold is unnecessary and can be removed. The following highlights are the same. %Authors: Could be removed (alredy did so).

\begin{figure}[H]
%\begin{center}
\includegraphics[width=0.96\textwidth]{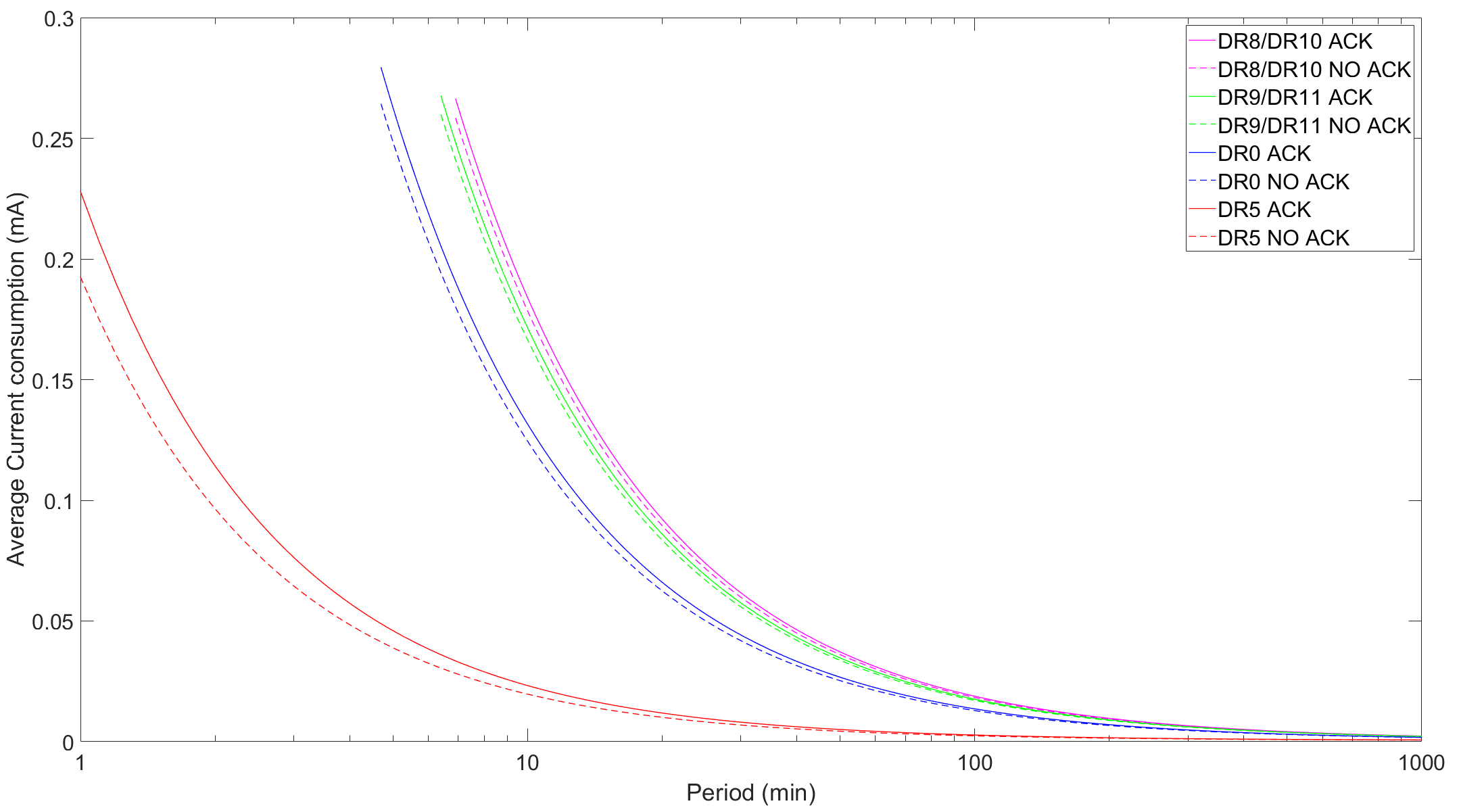}\vspace{-3pt}
\caption{ED average current consumption as a function of $T_{Period}$ for~DR8/DR10, DR9/DR11, DR0, and DR5 with~the maximum FRM Payload size permitted for each DR for both confirmed and unconfirmed~modes.}
\label{fig:avgCuACKvsNoACKMaxPayl}
%\end{center}
\end{figure}

Figure~\ref{fig:avgCuACKvsNoACKMaxPayl} shows the ED's average current consumption for different DRs, the~maximum allowed FRM Payload size for each DR, and~for both unconfirmed and confirmed transmission. As~expected, regardless of the DR, the~current consumption tends to converge towards the sleep state consumption of~around 500 nA as~the interval between transmissions (labeled “Period” in Figure~\ref{fig:avgCuACKvsNoACKMaxPayl} and subsequent figures) increases.

As expected, the~average current consumption for confirmed transmission is consistently higher than that of unconfirmed transmission. Specifically, the~differences between the confirmed and unconfirmed operational modes for the LR-FHSS DRs (i.e., for~the DR8/DR10 and DR9/DR11 pairs) are around 3\%. However, these differences rise to 5.7\% and 18.4\% for DR0 and DR5, respectively. Note that confirmed transmission can reduce the energy consumption as the probability of receiving the confirmation in the first receive window, $p_1$ (cf. Equation~(\ref{eq:IavgACK})), increases. The~results show the tendency of faster DRs, which have a shorter time on air (ToA), to reduce current consumption by a factor of up to around 10. Nonetheless, this pattern does not apply to DR9 and DR11, as~these DRs exhibit higher current consumption than DR0 although~having a higher data rate. This~oddity can be explained by the fact that the maximum FRM Payload size for DR9/DR11 (115 bytes) is substantially larger than that of DR0 (51 bytes) or DR8/DR10 (50 bytes), resulting in a %Please ensure meaning has been retained.
noticeably longer ToA for uplink transmission with DR9/DR11.

\begin{figure}[H]
\includegraphics[width=0.96\textwidth]{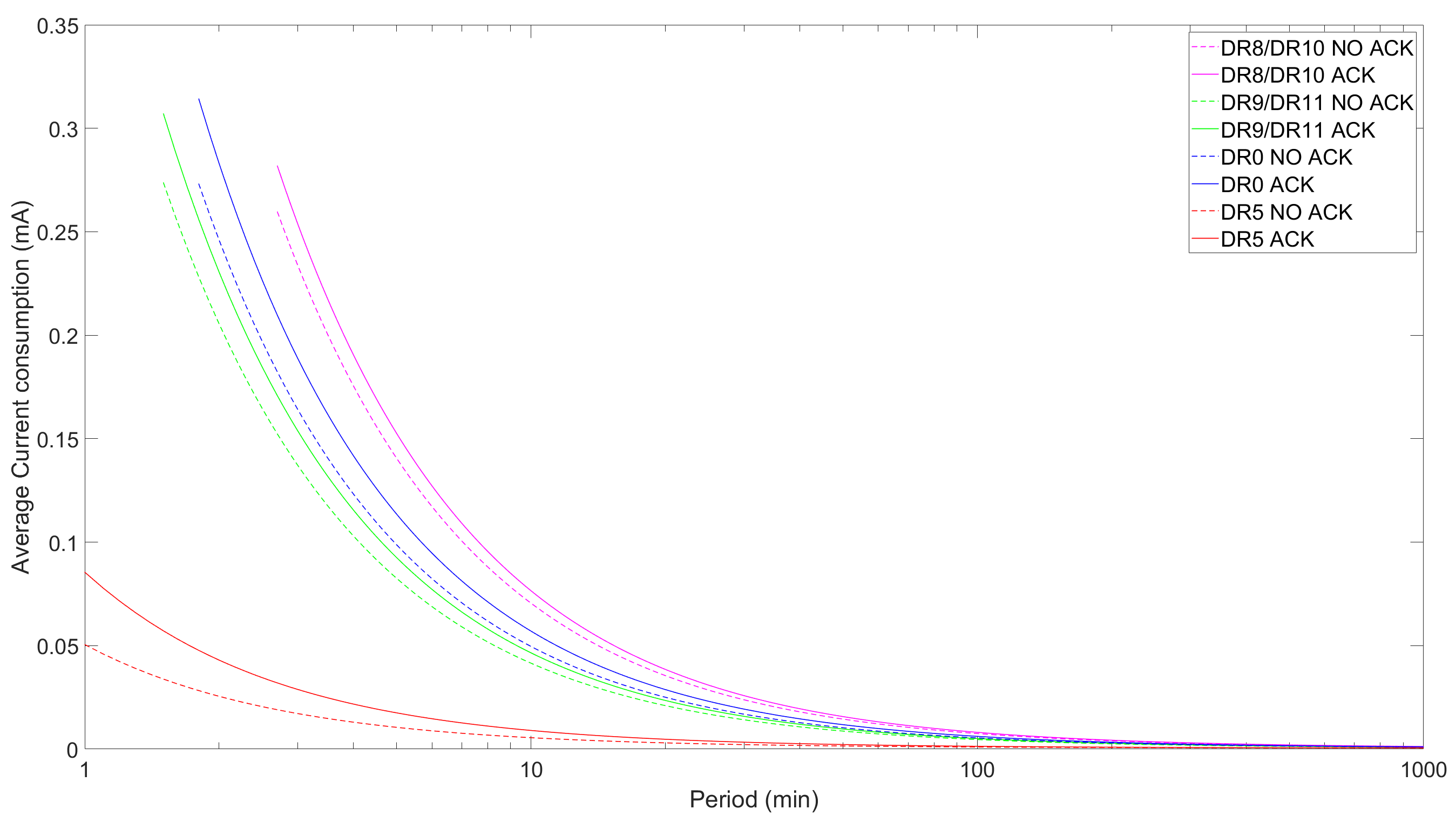}\vspace{-3pt}
\caption{ED average current consumption as a function of $T_{Period}$ for~DR8/DR10, DR9/DR11, DR0, and DR5 comparing confirmed and unconfirmed transmission for a 1-byte FRM Payload~size.}
\label{fig:avgCuACKvsNoACKMinPayl}
\end{figure}

However, when the payload size is the same for all DRs considered (e.g., as~shown in Table~\ref{tab:statesDRnoACK} for~a 1-byte FRM Payload), the~$T_{Tx}$ values correspond, in~descending order, to~DR8/DR10, DR0, DR9/DR11, and~DR5. This behavior is clearly illustrated in Figure~\ref{fig:avgCuACKvsNoACKMinPayl}, where average current consumption follows the same pattern. On~the other hand, there is a non-negligible difference between confirmed and unconfirmed transmission, with the former being more energy-consuming, especially for faster~DRs.

\unskip

\subsection{Battery~Lifetime}

The battery lifetime of an LR-FHSS ED is a crucial performance parameter in IoT networks as~it directly impacts the operational efficiency and maintenance requirements of devices that may have to be deployed in hard-to-reach locations. We evaluate ED battery lifetime by using Equations~(\ref{eq:Tlifetime}) and (\ref{eq:TlifetimeACK}). We assume a battery capacity of 230 mAh, which is typical for button cell batteries~\cite{varta_cr2032_farnell}. It is essential to remind the reader that our study solely accounts for the radio module's current usage. Consequently, the~increased current consumption caused by other ED tasks, such as internal communications, CPU activity, data processing, etc., will result in a decreased actual battery~lifespan.

\begin{figure}[H]
\includegraphics[width=0.96\textwidth]{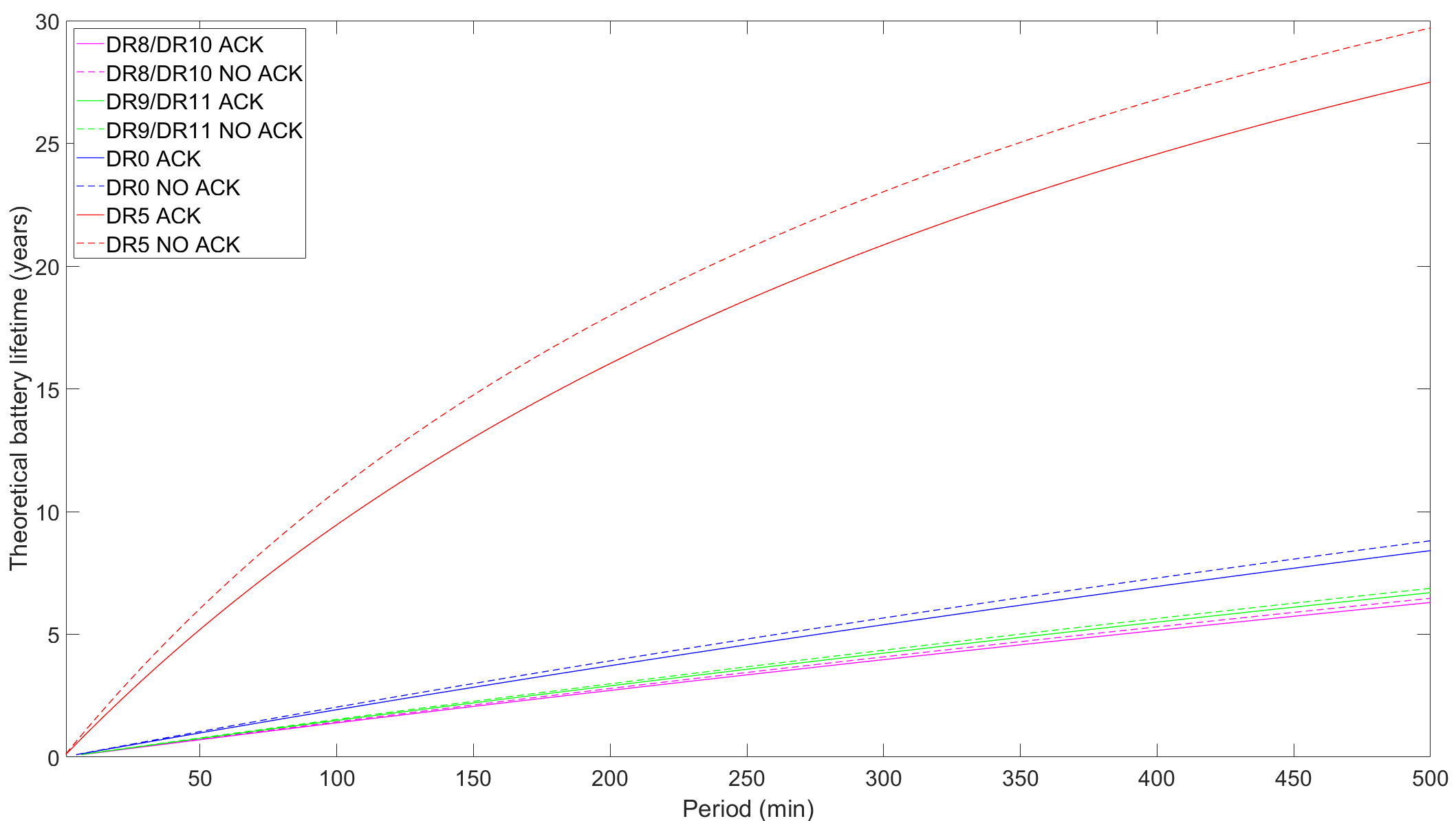}
\caption{Theoretical ED battery lifetime as a function of $T_{Period}$ for~DR8/DR10, DR9/DR11, DR0, and DR5 for~the maximum FRM Payload size permitted for each DR and comparing unconfirmed with confirmed uplink~transmission.}
\label{fig:bLifeACKVsNOACKMaxPayl}
\end{figure}

\begin{figure}[H]
\includegraphics[width=0.98\textwidth]{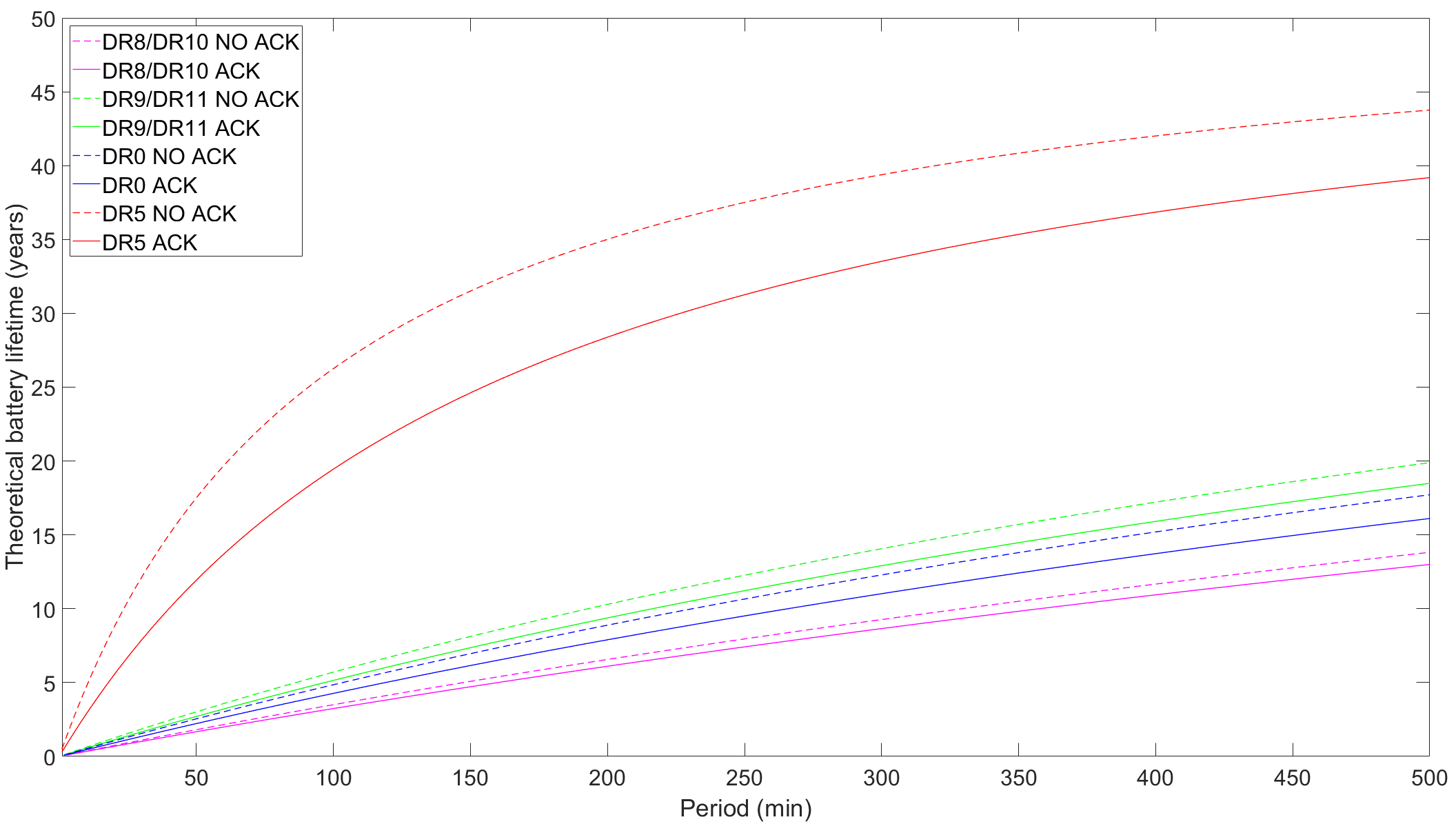}\vspace{-3pt}
\caption{Theoretical ED battery lifetime as a function of $T_{Period}$ for~DR8/DR10, DR9/DR11, DR0, and~DR5 with~1-byte FRM Payload transmissions and comparing unconfirmed and confirmed uplink~transmission.}
\label{fig:bLifeACKvsNoACKMinPayl}
\end{figure}

\begin{figure}[H]
\includegraphics[width=0.98\textwidth]{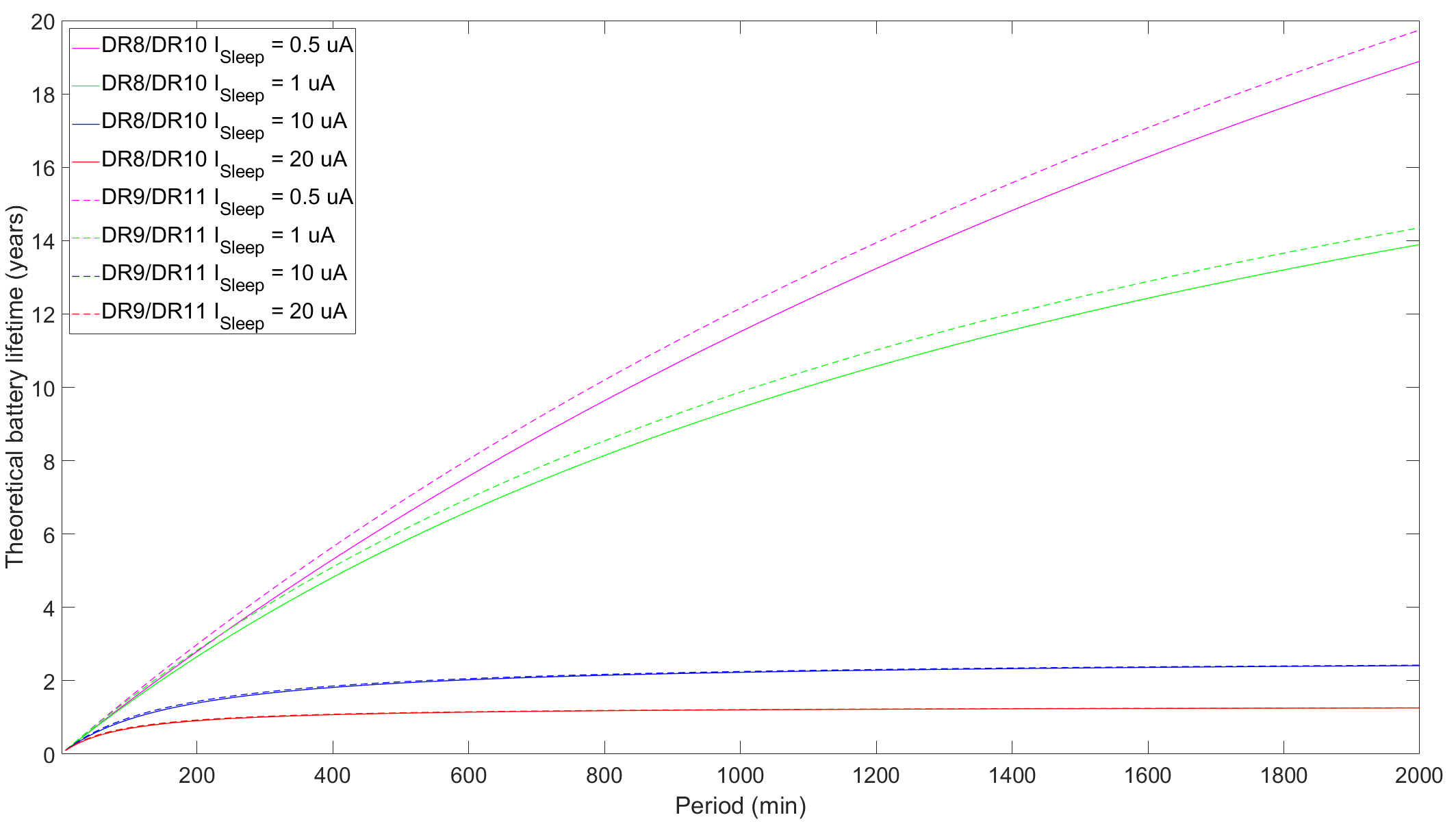}\vspace{-3pt}
\caption{Theoretical battery lifetime of the ED as a function of $T_{Period}$ for~DR8/DR10 and DR9/DR11 for~unconfirmed transmission, the~maximum FRM Payload size possible, and~different sleep current consumption~values.}
\label{fig:bLifeACKVsNOACKsleep}
\end{figure}

Figures~\ref{fig:bLifeACKVsNOACKMaxPayl}--\ref{fig:bLifeACKVsNOACKsleep} illustrate the theoretical ED battery lifetime as a function of $T_{Period}$ for~the maximum payload size allowed for each DR and~for a 1-byte payload, respectively. Consistent with the average current consumption study, the~ED battery lifetime increases asymptotically with $T_{Period}$. The~impact of using confirmed and unconfirmed transmission are shown in Figure~\ref{fig:bLifeACKVsNOACKMaxPayl}. For~$T_{Period}$ = 500 min, the battery lifetime for DR8/DR10 reaches around 6.5 years, while DR9/DR11 provide a slight enhancement of up to 6.9 years. If~we extend the measurement out of the scope of the figure, for~a one-day notification interval, the~theoretical battery lifetime for DR9/DR11 reaches around 16 years, while it slightly decreases to \textasciitilde15 years for DR8/DR10.

Except for DR5, battery lifetime differences between confirmed and unconfirmed transmission for~any of the DRs considered in Figure~\ref{fig:bLifeACKVsNOACKMaxPayl} are low. However, when $T_{Period}$ increases, these differences tend to slightly increase, yielding longer battery lifespans for unconfirmed transmission. Quantitatively, the~differences between confirmed and unconfirmed transmission for LR-FHSS DRs and DR0 are minor, with~the former showing differences of around 0.23 years. Nonetheless, for~DR5, the difference is notorious, reaching a peak of 2.2 years, which translates into an 8\% lifetime increase for unconfirmed transmission. Additionally, the results show greater differences in the lifetimes between confirmed and unconfirmed transmission for the DR0 and DR5 cases compared to the LR-FHSS ones. This is due to the varying contributions that different DRs set in the receive windows make to energy consumption, especially when measured relative to the $T_{Tx}$ values. For~example, for~a similar ToA in the uplink when comparing DR8/10, DR9/11, and~DR0, the~difference is more pronounced in the latter case, as~it uses a DR in the downlink that is the same one used in the uplink, which is the slowest~one.

DR5 lifetime results are significantly greater than those of the other DRs studied by a considerable margin due to its increased data rate (and lower ToA), which is in line with the findings in the average current consumption study. However, its reliability and communication range is also the worst among the considered DRs~\cite{semtech2021lrfhss}.

We also investigate the effect on battery lifetime of transmitting a modest 1-byte FRM Payload size (see Figure~\ref{fig:bLifeACKvsNoACKMinPayl}). In~particular, DR0 and the LR-FHSS DRs share similar results, as~the data rates are similar. DR8/DR10 are the most energy-consuming DRs in that group, while DR9/DR11 are the least consuming ones, with~theoretical battery lifetime results of up to around 20 years for a 500 min period. Then, DR5 outscores all the other DRs, as~it has the fastest bit rate and, consequently, the~lowest ToA. Also, the~use of ACKs reduces the theoretical battery lifetime, as~the ED has to decode the downlink frame and~one of the two receive windows is extended, which is especially impactful for DR5. Conclusively, the~DR used significantly affects the performance, with~the use of ACKs also contributing a non-negligible part. In~this regard, differences between confirmed and unconfirmed transmission range from 6.4 to 7.5\% for DR8/DR10 and DR9/DR11, respectively, up~to 11.74 \% (or 4.6 years) for DR5. This is due to the fact that as~the $T_{Tx}$ is lower, the~contribution of the ACK reception, which stays the same, is higher in relative~terms.

\vspace{-9pt}

As previously mentioned, the current consumption of a commercially ready ED, as opposed to that of a development kit, would be higher than in our case, as~we directly measured the current consumption of the radio module. For~this reason, we investigate how higher current consumption (i.e., emulating that of a whole device) during sleep stages would affect the battery lifetime performance (see Figure~\ref{fig:bLifeACKVsNOACKsleep}). We consider current consumption values of $0.5 \ \upmu$A (the one measured in our model), $1 \ \upmu$A, $10 \ \upmu$A, and~$20 \ \upmu$A (the last as assumed by Semtech, the~LR-FHSS chip manufacturer, in~\cite{semtech2021lrfhss} to be a typical LR-FHSS ED consumption in sleep mode). As~shown in the results depicted in Figure~\ref{fig:bLifeACKVsNOACKsleep}, battery lifetime stabilizes at around one and two years for $T_{Period}$ of around 5 h assuming sleep current, $I_{Sleep}$, values of $20 \ \upmu$A and $10 \ \upmu$A, respectively. On~the other hand, the~other $I_{Sleep}$ values show better battery lifetimes: for example, 8.5 years for $1\ \upmu$A and 10 years for $0.5 \ \upmu$A for a 6 h $T_{Period}$. As~expected, higher current consumption during the sleep state drastically shortens battery life, as~it is the most impactful state for higher $T_{Period}$ values. Particularly, this leads to up to a fifteen- to sixteen-fold reduction in battery life between the two most distant values (i.e., DR9/DR11 $I_{Sleep} = 0.5 \ \upmu$A and $20 \ \upmu$A).

\subsection{Energy~Cost}

The energy cost to transmit one bit of user data is the last performance parameter that we examine. We use Equations~(\ref{eq:energeticCostunACK}) and (\ref{eq:energeticCostACK}) to determine the energy cost of both unconfirmed and confirmed transmissions, respectively. %Please ensure meaning has been retained.
We use 3.3 V as the supply voltage of the device~\cite{semtech_lr1121}.

\begin{figure}[H]
\includegraphics[width=1\textwidth]{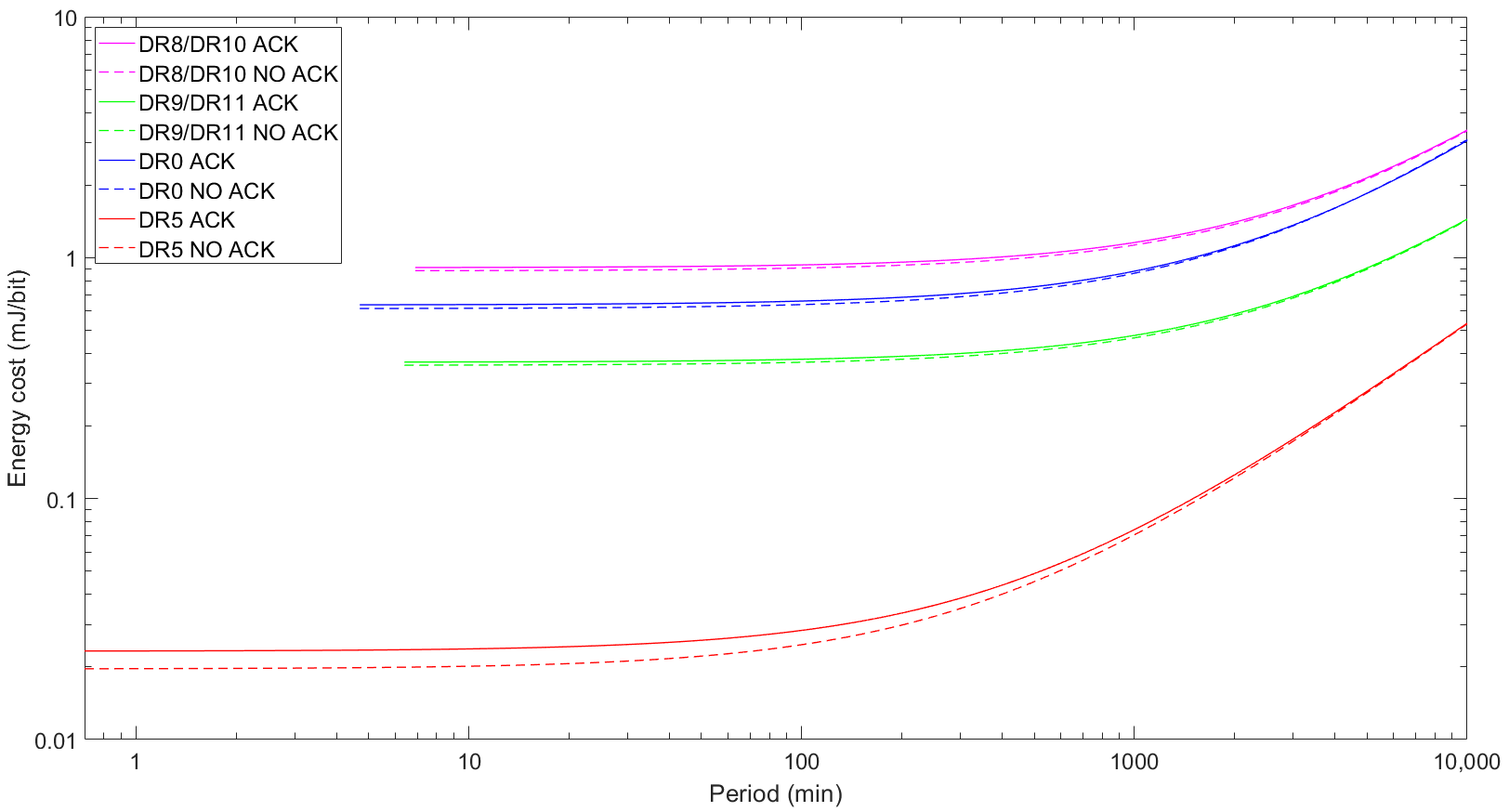}
\caption{Energy cost of unconfirmed and confirmed data transmission as a function of $T_{Period}$ for~the maximum FRM Payload sizes for DR8/DR10, DR9/DR11, DR0, and~DR5.}
\label{fig:enCoNoACKvsACKMaxPay}
\end{figure}

The energy cost of sending the maximum FRM Payload for every DR with and without ACK is displayed in Figure~\ref{fig:enCoNoACKvsACKMaxPay}. Note that for~a given DR, the~average current consumption tends to the sleep state current consumption ($I_{Sleep}$) when $T_{Period}$ grows, but~the number of transmitted data bits stays constant. Therefore, longer $T_{Period}$ results in a greater energy cost, as~the energy consumption grows for the same amount of transmitted data. DR9/DR11 outperform DR8/DR10 in terms of energy cost, despite performing similarly in terms of average current consumption, for~maximum FRM Payloads. The~payload sizes play a key role in this phenomenon: DR9 and DR11 can deliver more data at a time and, consequently, achieve a lower energy cost, since their maximum FRM Payload size is 115 bytes, while DR8 and DR10 can only carry up to 50 bytes for their maximum FRM Payload. However, DR9 and DR11 are still much less efficient than DR5 (although they even outperform DR0 in terms of energy efficiency). Due to its higher bit rate, shorter ToA, and~higher maximum FRM Payload size, DR5 is the most energy efficient. For~a given DR, the~energy cost difference between confirmed and unconfirmed transmission is not negligible, but~it follows a very similar tendency across DRs, where the former is slightly less~efficient.

\begin{figure}[H]
\includegraphics[width=1\textwidth]{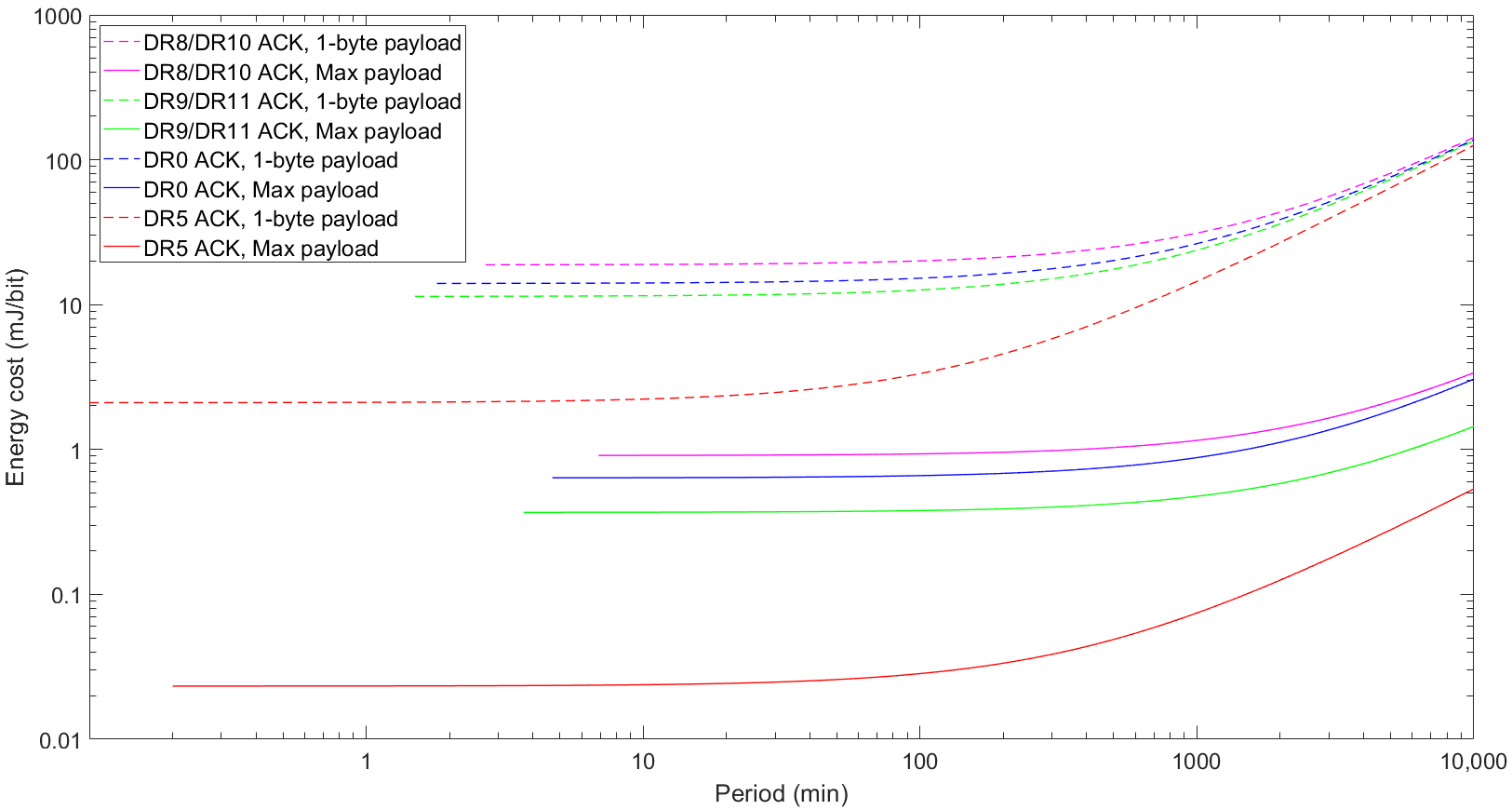}
\caption{Energy cost of confirmed data transmission as a function of $T_{Period}$ when sending the maximum allowable FRM Payload and a 1-byte FRM Payload for~DR8/DR10, DR9/DR11, DR0, and~DR5.}
\label{fig:enCoACKMaxVsMinPay}
\end{figure}

The energy cost differences between sending a 1-byte FRM Payload size and the maximum FRM Payload size are shown in Figure~\ref{fig:enCoACKMaxVsMinPay}. Maximizing the FRM Payload size has a significant impact on the energy cost of data transmission, as~the energy consumed per every transmitted bit of data is lower. This behavior also depends on the DR used because~greater-bit-rate-capable DRs like DR5 or DR9/11 achieve the best results. Note that the combination of bit rates and higher maximum FRM Payload sizes makes these DRs the most energy-efficient ones. Nonetheless, it should be noted that DR5 outperforms all other studied DRs, achieving nearly two orders of magnitude better performance for the shortest $T_{Period}$ values. Regarding DR8/DR10, the~energy cost decreases up to around 21~times when the payload is maximized, and~the decrease is %Please ensure meaning has been retained.
around 31 times for the DR9/DR11~case.

%MDPI: Please use commas to separate thousands for numbers with five or more digits (not four digits) in the picture, e.g., “10000” should be “10,000”. %Authors: Done
\vspace{-9pt}
%MDPI: Please use commas to separate thousands for numbers with five or more digits (not four digits) in the picture, e.g., “10000” should be “10,000”. %Authors: Done

%%%%%%%%%%%%%%%%%%%%%%%%%%%%%%%%%%%%%%%%%%
\section{Conclusions}\label{sec:conclusions}

In this article, we provided the first energy consumption model of a Class A LoRaWAN ED using the novel LR-FHSS physical layer. We also evaluated the impact of variables like the use of confirmed transmissions, DRs, and FRM Payload sizes. The~key metrics considered were the average current consumption, battery lifetime, and~energy cost of data transmission.
The model was created based on data collected using a real LoRaWAN network in which all of the components support~LR-FHSS.

From the evaluation of the model, a~first general conclusion is that LR-FHSS DRs exhibit energy performance comparable to that of the most robust LoRa DR (i.e., DR0) but~worse than a fast LoRa DR such as~DR5.

The average current consumption for DR8/DR10 was consistently the highest in both the maximum and 1-byte FRM Payload cases, reaching a peak of around 0.27 mA. The~average current consumption for DR9/DR11 behaved very similarly to the former when using the maximum FRM Payload size, although~it decreased when using a 1-byte FRM Payload. For~a constant FRM Payload size, the~average current consumption decreased with the physical-layer bit rate due to the impact of the latter on ToA. DR5 achieved up to approximately six-times lower current consumption. Regarding the use of ACKs, the~ED consumed less current when using the unconfirmed approach (e.g., around 3\% lower for all LR-FHSS DRs). Overall, the~current consumption tended to decrease with the period between uplink transmissions, since the influence of sleep current increased as~well.

The theoretical battery lifetime of an ED with a 230 mAh button-like battery is \textasciitilde20~years for a transmission period of 500 min, unconfirmed transmission, DR9/DR11, and~a 1-byte payload. Under~the same conditions, DR8/DR10 reaches a maximum of around \mbox{14 years}, DR0 yields an intermediate lifetime of \textasciitilde18 years, and DR5 is the best-performing among the considered DRs. When maximizing the payload in an infrequent interval, such as a one-day period, the~battery lifetime for DR9/DR11 is 16 years. As~expected, battery lifetime decreases with payload size and sleep current consumption. Nonetheless, although~DR9/DR11 uses a faster bit rate than DR8/DR10, their lifetimes are similar when using the maximum FRM Payload permitted by each DR. The~outcome is that maximizing the bit rate, minimizing the payload size, and not using ACKs enhances the lifetime of~EDs.

Finally, the~energy cost per transmitted bit is lower for DRs that feature a higher bit rate and a larger payload size. For~example, the~energy cost of DR9/DR11 is lower than that of DR8/DR10, with~differences of up to around 2.5 times. Not using ACKs is more energy efficient regardless of the DR, but~by a mild margin. It should also be noted that the energy per bit tends to increase as the period between uplink transmissions increases due to the higher total energy consumption for the same amount of transmitted~data.

%%%%%%%%%%%%%%%%%%%%%%%%%%%%%%%%%%%%%%%%%%
%\section{Patents}

%This section is not mandatory, but may be added if there are patents resulting from the work reported in this manuscript.

%%%%%%%%%%%%%%%%%%%%%%%%%%%%%%%%%%%%%%%%%%
\vspace{6pt} 

%%%%%%%%%%%%%%%%%%%%%%%%%%%%%%%%%%%%%%%%%%
%% optional
%\supplementary{The following supporting information can be downloaded atab:  \linksupplementary{s1}, Figure S1: title; Table S1: title; Video S1: title.}

% \supplementary{The following supporting information can be downloaded atab: \linksupplementary{s1}, Figure S1: title; Table S1: title; Video S1: title. A supporting video article is available at doi: link.}

% Only for journal Hardwareq:
% \supplementary{The following supporting information can be downloaded atab: \linksupplementary{s1}, Figure S1: title; Table S1: title; Video S1: title.\vspace{6pt}\\
%\begin{tabularx}{\textwidth}{lll}
%\toprule
%\textbf{Name} & \textbf{Type} & \textbf{Description} \\
%\midrule
%S1 & Python script (.py) & Script of python source code used in XX \\
%S2 & Text (.txt) & Script of modelling code used to make Figure X \\
%S3 & Text (.txt) & Raw data from experiment X \\
%S4 & Video (.mp4) & Video demonstrating the hardware in use \\
%\bottomrule
%\end{tabularx}
%}

%%%%%%%%%%%%%%%%%%%%%%%%%%%%%%%%%%%%%%%%%%
\vspace{6pt}
\authorcontributions{Conceptualization, R.S.-V., C.G. and E.G.-V.; methodology, R.S.-V. and \mbox{B.H.-S}.; software, R.S.-V., L.C. and B.H.-S.; validation, R.S.-V., L.C. and R.V.; formal analysis, R.S.-V.; investigation, R.S.-V., L.C., B.H.-S. and C.G.; resources, L.C. and R.V.; data curation, R.S.-V., B.H.-S. and R.V.; writing—original draft preparation, R.S.-V. and B.H.-S.; writing—review and editing, R.S.-V., L.C., R.V., C.G. and E.G.-V.; visualization, R.S.-V. and B.H.-S.; supervision, L.C., C.G. and E.G.-V.; project administration, C.G. and E.G.-V.; funding acquisition, E.G.-V. All authors have read and agreed to the published version of the~manuscript.}

\funding{This research was funded %MDPI: Information regarding the funder and the funding number should be provided. Please check the accuracy of funding data and any other information carefully. Any updates after publication should be carefully considered. %Authors: The funding information is correct.
\textls[-15]{by the Spanish Government MCIN/AEI/10.13039/501100011033 through projects PID2019-106808RA-I00 and PID2023-146378NB-I00 and~by the Secretaria d'Universitats} i Recerca del Departament d'Empresa i Coneixement de la Generalitat de Catalunya 2021 through grant SGR 00330. The~first author gratefully acknowledges a predoctoral program AGAUR-FI grant (2023 FI-1 00154) Joan Oró of the Secretaria d'Universitats i Recerca del Departament de Recerca i Universitats de la Generalitat de Catalunya and the European Fund Social Plus.}

\institutionalreview{Not applicable.}

\informedconsent{Not applicable.}

\dataavailability{The original contributions presented in the study are included in the article; further inquiries can be directed to the corresponding author.}

\acknowledgments{The authors would like to thank the Semtech team for kindly providing the gateway firmware update to enable LR-FHSS, as~well as for their subsequent~support.}

\conflictsofinterest{The authors declare no conflicts of~interest.}

%%%%%%%%%%%%%%%%%%%%%%%%%%%%%%%%%%%%%%%%%%
%% Optional

%% Only for journal Encyclopedia
%\entrylink{The Link to this entry published on the encyclopedia platform.}

\abbreviations{Abbreviations}{
The following abbreviations are used in this manuscript:\\

\noindent 
\begin{tabular}{@{}ll}
ACK & Acknowledgment\\
CAD & Channel Activity Detection\\
CR & Coding Rate\\
DFH & Dynamic Frequency Hopping\\
DR & Data Rate\\
DtS-IoT & Direct-to-Satellite IoT\\
ED & End Device\\
FEC & Forward Error Correction\\
FHS & Frequency Hopping Sequence\\
FSK & Frequency Shift Keying\\
GW & Gateway\\
IoT & Internet of Things\\
LoRa & Long Range\\
LoRaWAN & Long Range Wide Area Network\\
LR-FHSS & Long Range - Frequency Hopping Spread Spectrum\\
LPWAN & Low Power Wide Area Network\\
MAC & Media Access Control\\
NS & Network Server\\
PHY & Physical Layer\\
ToA & Time on Air
\end{tabular}
}

%%%%%%%%%%%%%%%%%%%%%%%%%%%%%%%%%%%%%%%%%%
%% Optional (Appendix)
\appendixtitles{no} % Leave argument "no" if all appendix headings stay EMPTY (then no dot is printed after "Appendix A"). If~the appendix sections contain a heading then change the argument to "yes".
\appendixstart
\appendix
\section[\appendixname~\thesection]{}\label{sec:appendix}
%\subsection[\appendixname~\thesubsection]{}
This appendix provides the original (and~inaccurate) expression for $T_{payload}$ that is offered in the LoRaWAN Regional Parameters v1.0.4 specification from the LoRa Alliance~\cite{regional_parameters_v104} and~is also used in related work. Equation~(\ref{eq:tpayloadappendix}) calculates $T_{payload}$ as explicitly indicated in the document.
\begin{equation}\label{eq:tpayloadappendix}
T_{payload} = \begin{cases}
\left \lceil \frac{L_{PHY}+3}{2} \right \rceil \times 102.4 \; ms & \text{if } CR = 1/3, \\[0.3cm]
\left \lceil \frac{L_{PHY}+3}{4} \right \rceil \times 102.4 \; ms  & \text{if } CR = 2/3.
\end{cases}
\end{equation}

%%%%%%%%%%%%%%%%%%%%%%%%%%%%%%%%%%%%%%%%%%
\begin{adjustwidth}{-\extralength}{0cm}

\reftitle{References}


\bibliography{bibliography.bib}

\PublishersNote{}
\end{adjustwidth}
\end{document}